\documentclass[12pt]{amsart}

\usepackage[T1]{fontenc}
\usepackage[utf8]{inputenc}
\usepackage{amsmath,amsthm,amssymb,amsfonts}
\usepackage{enumerate}
\usepackage{mathrsfs}
\usepackage{hyperref}
\usepackage{tikz}
\usetikzlibrary{calc, arrows.meta}
\usepackage{microtype}
\usepackage{array}

\allowdisplaybreaks
\hypersetup{colorlinks=true,linkcolor=blue,citecolor=blue,urlcolor=blue}

\newcommand{\Z}{\mathbb Z}
\newcommand{\Q}{\mathbb Q}
\newcommand{\R}{\mathbb R}
\newcommand{\C}{\mathbb C}
\newcommand{\Zp}{\mathbb Z_p}
\newcommand{\Qp}{\mathbb Q_p}
\newcommand{\Cp}{\mathbb C_p}
\newcommand{\dd}{\,\mathrm d}

\newcommand{\Ep}{\mathcal{E}_p}
\newcommand{\nT}[1]{\|#1\|_{\mathrm{Tate}}}
\newcommand{\sinp}{\operatorname{sin}_p}
\newcommand{\cosp}{\operatorname{cos}_p}
\newcommand{\rhop}{p^{-\frac{1}{p-1}}}
\newcommand{\Sfree}{S_{\mathrm{free}}}

\newcommand{\sK}{\mathcal{K}}
\newcommand{\Kp}{\mathcal{K}_p}
\newcommand{\Kkin}{\mathcal{K}_p^{\mathrm{kin}}}
\newcommand{\Kvar}{\mathcal{K}_p^{\mathrm{var}}}
\newcommand{\Dp}{\mathcal{D}_p}
\newcommand{\Rp}{\mathcal{R}_p}
\newcommand{\sT}{\mathcal T}

\newcommand{\sA}{\mathscr A}
\newcommand{\sE}{\mathscr E}
\newcommand{\sX}{\mathscr X}

\newtheorem{theorem}{Theorem}[section]
\newtheorem{lemma}[theorem]{Lemma}
\newtheorem{proposition}[theorem]{Proposition}
\newtheorem{corollary}[theorem]{Corollary}

\theoremstyle{definition}
\newtheorem{definition}[theorem]{Definition}
\newtheorem{remark}[theorem]{Remark}
\newtheorem{example}[theorem]{Example}
\newtheorem{convention}[theorem]{Convention}

\numberwithin{equation}{section}
\begin{document}

\title[Block decimation for $p$-adic valued path integrals]
{Block decimation and perfect actions\\ for $p$-adic valued path integrals}

\author{Su Hu}
\address{Department of Mathematics, South China University of Technology,
Guangzhou, Guangdong 510640, China}
\email{mahusu@scut.edu.cn}
\author{Min-Soo Kim}
\address{Department of Mathematics Education, Kyungnam University,
Changwon, Gyeongnam 51767, Republic of Korea}
\email{mskim@kyungnam.ac.kr}


\begin{abstract}
In earlier work we introduced a finite-partition propagator for
$\mathbb{C}_p$-valued wave functions, in which the intermediate positions
are integrated against atomic measures. We show that this iterated integral
evaluates the action on the equally spaced chain between the endpoints, so
that each partition level is obtained by iterating an exact, tree-level
block decimation of two-point lattice actions: no fluctuations are
integrated out, and no multi-point couplings or prefactors are generated.
This is why the free propagator does not depend on the partition. This
kinematic decimation is the Fr\'echet derivative, at the free actions, of
the variational (stationary-point) decimation used to construct classical
perfect actions. Since the time step $p^n$ is small in the real norm
exactly when it is large $p$-adically, refinement and coarsening (the
ultraviolet and infrared directions over $\mathbb{R}$) exchange their
roles, and each yields one convergent limit. Under refinement, for
potentials supported on $\mathbb{Z}_p$, given there by a convergent power
series, and entering the slice action without the time-step weight, the
interaction stabilizes off the diagonal after finitely many steps and
converges uniformly on $\mathbb{Z}_p^2$, with sharp error $p^{-M}$ after
$M$ steps, to a scale-covariant (perfect) action. The limit is the unique
scale-covariant action that agrees with the bare potential at large
separation, and near the diagonal it is governed by a Volkenborn integral;
with the time-step weight no limit exists. For the harmonic oscillator no
refinement limit exists at fixed frequency, but under coarsening the
variational flow is solved in closed form: the lattice angle is multiplied
by $p$ at each step, the coupling obeys a Chebyshev recursion, no caustic
occurs, and the rescaled action converges to a free action with
renormalized mass $m_0\sin_p(\theta_0)/\theta_0$, where $\theta_0$ is the
initial lattice angle. The rescaled linearized map is diagonalizable on
polynomial actions with eigenvalues $p^{2-b}$, $b\ge 0$, so that directions
that are relevant over $\mathbb{R}$ are irrelevant over $\mathbb{Q}_p$ and
conversely.
\end{abstract}
\maketitle

\section{Introduction}\label{sec:intro}

\subsection{Background}

Throughout, $p$ is a fixed odd prime, $\Cp$ is the completion of an algebraic closure of $\Qp$,
$v_p$ is the valuation on $\Cp$ with $v_p(p)=1$, $|x|_p=p^{-v_p(x)}$,
$\Zp=\{x\in\Qp:|x|_p\le1\}$, and
\[
  \Ep:=\big\{z\in\Cp : |z|_p<\rhop\big\}
\]
is the disc of convergence of the $p$-adic exponential.

Following the proposal of Volovich \cite{Volovich1987} and Vladimirov--Volovich \cite{VV1989} that
geometry at very small scales may be non-Archimedean (see \cite{Dragovich2009} for a survey),
quantum mechanics over $\Qp$ has developed in two directions. In the first
\cite{VV1989,VVZ1994,Zelenov1991}, wave functions are complex-valued, $\psi:\Qp\to\C$. Haar measure
and $L^2$ spectral theory are then available (see \cite{ZunigaGalindo2024} for recent work), path
integrals for quadratic actions have been computed \cite{DjordjevicDragovich1997}, and the
renormalization group has a long history: Dyson's hierarchical model \cite{Dyson1969} and its
analysis by Bleher and Sinai \cite{BleherSinai1973} are models over $\Qp$ \cite{LernerMissarov1989},
rigorous renormalization-group constructions exist for $p$-adic field theories \cite{ACG2013}, and
the Bruhat--Tits tree serves as a bulk geometry in $p$-adic AdS/CFT \cite{Gubser2017}. In the second
direction, initiated by Khrennikov \cite{Khrennikov1991,Khrennikov1994book,AK1996}, the wave
functions themselves take values in a non-Archimedean field, $\psi:\Qp\to\Cp$ or $\Cp\to\Cp$. Two
basic tools are then lost. There is no non-zero bounded translation-invariant $\Cp$-valued measure
on $\Qp$ \cite{vanRooij1978,Schikhof1984}, and there is no spectral theory of a Laplacian
comparable to the $L^2$ theory \cite{Vishik}. Time-sliced actions and finite-partition path
integrals therefore become the primary objects.

\subsection{The construction}

In \cite{HuKim2026}, referred to below as HK, we defined a finite-partition propagator in the
second setting. A $p$-adic time segment is divided into $p^{N-n}$ subsegments by Zelenov's ball
partition at level $n$, the action is summed over the slices, and the intermediate positions are
integrated against atomic measures whose centres depend on the neighbouring positions. For the free
particle the result does not depend on $n$, so no limit is needed. Two questions were left open:
why the dependence on $n$ cancels, and what happens under refinement when a potential is present.
This paper answers both.

The starting point is an observation about what the HK integral computes. Integration against an atomic measure is point evaluation, so the iterated integral evaluates the total action on a single
configuration, the equally spaced chain between the endpoints (Lemma~\ref{lem:nested}). There is no
integration over fluctuations and no determinant. Passing from level $n$ to level $n+1$ is
therefore the map
\begin{equation}\label{eq:intro_decim}
  S\ \longmapsto\ \Kp S,\qquad
  (\Kp S)(x,y)=\sum_{r=0}^{p-1}S\big(\gamma_r(x,y),\gamma_{r+1}(x,y)\big),
\end{equation}
on two-point actions, where $\gamma_0=x,\gamma_1,\dots,\gamma_p=y$ are the nodes of a block of $p$
slices. In the language of lattice field theory, \eqref{eq:intro_decim} is a blocking
transformation at tree level. The free actions $\Sfree(x,y;\epsilon)=m(y-x)^2/(2\epsilon)$ satisfy
$\Kp\Sfree(\cdot,\cdot;\epsilon)=\Sfree(\cdot,\cdot;p\epsilon)$, and this is the reason for the
cancellation observed in HK.

We study two versions of \eqref{eq:intro_decim}. In the \emph{kinematic} decimation $\Kkin$ the
nodes are equally spaced, as in HK. In the \emph{variational} decimation $\Kvar$ the interior nodes
are the stationary points of the block action; this is the classical perfect-action construction of
Hasenfratz and Niedermayer \cite{Hasenfratz1994}. The two rules agree on free actions, and $\Kkin$
is the Fr\'echet derivative of $\Kvar$ at a free action, with an explicit quadratic remainder
(Proposition~\ref{prop:tangent}). Statements about $\Kkin$ are thus exact statements about the HK
propagator and first-order statements about the variational flow.

\subsection{Main results}

\emph{Potentials supported on $\Zp$} (Section~\ref{sec:perfect}). Let $V$ vanish outside $\Zp$.
Under $\Kkin$, the interaction part $W_{N-n}$ of the level-$n$ effective action stops changing at
each point $(x,y)\in\Zp^2$ with $x\ne y$ once $N-n\ge v_p(y-x)$, because the new midpoints leave
the support of $V$ (Proposition~\ref{prop:FP}). The pointwise limit $W^*$ satisfies the covariance
identity $\Kkin S^*_\epsilon=S^*_{p\epsilon}$, and it is the only solution of this identity that
agrees with the bare potential at large separation (Proposition~\ref{prop:char}). If moreover $V$
is given on $\Zp$ by a power series with coefficients tending to zero, then as $n\to-\infty$ the
convergence is uniform on $\Zp^2$, with error at most $p^{-(N-n)}\nT{V}$, and this bound is
attained (Theorem~\ref{thm:uniform_RG}). Near the diagonal, $W^*(x,y)$ equals $|y-x|_p^{-1}$ times
a Volkenborn integral of $V$, up to an error of order $|y-x|_p^2$
(Proposition~\ref{prop:Volkenborn}). The rate comes from the Clausen--von Staudt bound on power
sums. It is essential that the potential enters the slice action without the factor $\epsilon_n$,
as in HK; with that factor the limit does not exist (Proposition~\ref{prop:epsweight}).

\emph{Harmonic oscillator} (Section~\ref{sec:oscillator}). For the oscillator, $\Kvar$ can be
iterated in closed form in the coarsening direction $\epsilon_\ell=p^{n_0+\ell}$ ($\ell\ge0$). The
$\ell$-th action is of Mehler type with angle $p^\ell\theta_0$ (Theorem~\ref{thm:oscillator}), and
since $\sinp$ has no zero in $\Ep\setminus\{0\}$, no caustic occurs at any step. The dimensionless
coupling satisfies $g_{\ell+1}=2(-1)^{(p-1)/2}T_p(g_\ell/2)$ with $T_p$ the Chebyshev polynomial,
and $|g_\ell|_p=p^{-\ell}|g_0|_p$. The rescaled action converges to the free action with mass
$m^*=m_0\sinp(\theta_0)/\theta_0$, and its distance from the limit in the supremum norm on $\Zp^2$
equals $|3|_p^{-1}|m_0|_p|g_0|_p^2p^{-2\ell}$. The algebra is that of the real time-sliced
oscillator \cite{Kleinert2009,MarsdenWest2001}; what is specific to the $p$-adic case is the
direction of the flow. A refinement limit at fixed frequency does not exist
(Proposition~\ref{prop:noUV}).

\emph{Linearized map} (Sections~\ref{sec:RGop} and~\ref{sec:spectrum}). The map $\sT=p\,\Kkin$ is
diagonalized on polynomial actions: the polynomials
$(v-u)^{b-1}\big(v^{k+1}-u^{k+1}\big)/(k+1)$ are eigenvectors with eigenvalue $p^{2-b}$
(Proposition~\ref{prop:spectrum}). Hence the potential directions ($b=0$) contract $p$-adically,
the kinetic direction ($b=2$) is marginal, and the higher-gradient directions ($b\ge4$) expand;
over $\R$ the roles of $b=0$ and $b\ge4$ are exchanged. In particular the rescaled map is unbounded
for every norm on a space containing all powers $(v-u)^b$ (Proposition~\ref{prop:nogo}), so the
convergence results above cannot come from a contraction principle and are proved directly.

\subsection{Comparison with the Archimedean renormalization group}

In Kadanoff's block-spin picture \cite{Kadanoff1966} and Wilson's formulation
\cite{Wilson1971,WilsonKogut1974}, a renormalization step integrates out short-distance variables
and rescales; see \cite{Efrati2014} for real-space methods. In lattice field theory a blocking
kernel couples fine variables to coarse ones, and a fixed point of the transformation is called a
perfect action \cite{Hasenfratz1994}. Table~\ref{tab:dict} compares that picture with the present
one. The inversions in the last three rows stem from one arithmetic identity:
$\epsilon_n=p^n\in\Q$, and $|p|_\infty|p|_p=1$.

\begin{table}[htbp]
\centering
\small
\caption{Archimedean lattice renormalization versus $p$-adic block decimation.}
\label{tab:dict}
\begin{tabular}{@{}>{\raggedright\arraybackslash}p{0.23\textwidth}>{\raggedright\arraybackslash}p{0.34\textwidth}>{\raggedright\arraybackslash}p{0.36\textwidth}@{}}
\hline
 & \textbf{Archimedean lattice} & \textbf{Present $p$-adic setting} \\ \hline
Block aggregation & Integral over block variables & Sum along a chain, \eqref{eq:Kadanoff_def}; no integration \\
Blocking kernel & Weighted block average & Dirac measures at free or stationary nodes \\
Generated couplings & Infinitely many; truncation & None: two-point actions map to two-point actions \\
Fluctuation prefactor & Van Vleck factor, caustics & None \\
Perfect action & Fixed point of the blocking map & Scale-covariant family, Definition~\ref{def:perfect} \\
Refinement $n\to-\infty$ & $|\epsilon_n|_\infty\to0$ & $|\epsilon_n|_p\to\infty$ \\
Coarsening $\ell\to+\infty$ & $|\epsilon_\ell|_\infty\to\infty$ & $|\epsilon_\ell|_p\to0$ \\
Eigenvalue $p^{2-b}$ & Relevant for $b=0$, irrelevant for $b\ge4$ & Contracting for $b=0$, expanding for $b\ge4$ \\ \hline
\end{tabular}
\end{table}

\subsection{Organization}
Section~\ref{sec:model} recalls the partition model, gives the iterated Dirac integral an explicit
inductive definition, and replaces the slice-wise admissibility condition of HK by a condition on
the total action. Section~\ref{sec:RGop} introduces the decimation operators and proves the
linearization theorem. Section~\ref{sec:perfect} treats potentials supported on $\Zp$,
Section~\ref{sec:oscillator} the harmonic oscillator, and Section~\ref{sec:spectrum} the spectrum
of the linearized map. Section~\ref{sec:discussion} lists open problems. The appendices contain
worked examples for $p=3$ and $p=5$ in exact rational arithmetic.

\section{The model}\label{sec:model}

\subsection{Partitions and the scale parameter}

Let $t\in\Qp$ with $|t|_p=p^{-N-1}$ for an integer $N\ge0$. Zelenov \cite{Zelenov1991} orders $\Qp$
by means of a continuous injection $\phi:\Qp\to\R$ and defines segments $[a,b]$ accordingly. The
segment $[0,t]$ lies in the disc $B_N=\{\tau\in\Qp:|\tau|_p<p^N\}$, which is the segment
$[0,-p^{-N}]$ with its right endpoint removed. For every integer $n\le N$, $B_N$ is the disjoint
union of $K=p^{N-n}$ discs of radius $p^n$. Their endpoints, listed in increasing order, are
$0=t_0<t_1<\dots<t_K=-p^{-N}$, and $|t_{i+1}-t_i|_p=p^n$ \cite{HuKim2026}. As in HK we set
\begin{equation}\label{eq:scale}
  \epsilon_n:=p^n\in\Q\subset\Cp,\qquad K\epsilon_n=p^N,\qquad |\epsilon_n|_p=p^{-n}.
\end{equation}
HK takes $n\le N-1$; we also admit $n=N$, that is, $K=1$.

\begin{remark}\label{rem:scale}
(i) The parameter $\epsilon_n$ is the real number $|t_{i+1}-t_i|_p$ regarded as an element of
$\Q\subset\Cp$; it is not the $p$-adic increment $t_{i+1}-t_i$.
(ii) Under refinement ($n\to-\infty$), $\epsilon_n\to0$ in $\R$ while $|\epsilon_n|_p\to\infty$.
This inversion is used throughout.
(iii) The partition covers $[0,-p^{-N}]$, so the propagator depends on $t$ only through $N$; in HK,
$p^N$ is identified with $p^{-1}|t|_p^{-1}$.
\end{remark}

\subsection{Atomic measures and characters}

For $d\in\Qp$ let $\delta_d(U):=\mathbf 1_U(d)$ on compact-open sets $U\subseteq\Qp$. This is a
bounded $\Cp$-valued measure in the sense of Monna--Springer integration
\cite{Schikhof1984,vanRooij1978}. In that theory one attaches to a measure $\mu$ the function
$N_\mu(x):=\inf_{U\ni x}\sup_{U'\subseteq U}|\mu(U')|_p$ ($U,U'$ compact-open) and the seminorm
$\|f\|_\mu:=\sup_x|f(x)|_pN_\mu(x)$, and the integrable functions are the $\|\cdot\|_\mu$-limits
of locally constant functions with compact support. For $\mu=\delta_d$ one has $N_\mu(d)=1$ and
$N_\mu(x)=0$ for $x\ne d$, so $\|f\|_{\delta_d}=|f(d)|_p$. Every function $f$ is therefore
integrable, with
\begin{equation}\label{eq:dirac}
  \int_{\Qp}f\dd\delta_d=f(d),
\end{equation}
and only the value of $f$ at $d$ matters. The measures $\delta_d$ are not translation invariant,
and this cannot be avoided: a translation-invariant measure $\mu$ satisfies
$\mu(\Zp)=p^{j}\mu(p^{j}\Zp)$ for $j\ge0$, so that $|\mu(p^j\Zp)|_p=p^j|\mu(\Zp)|_p$;
boundedness forces $\mu(\Zp)=0$, and likewise for every ball.

The $p$-adic exponential $e_p(z)=\sum_{j\ge0}z^j/j!$ converges on $\Ep$, where
$e_p(z_1+z_2)=e_p(z_1)e_p(z_2)$, $|e_p(z)|_p=1$ and $|e_p(z)-1|_p=|z|_p$
\cite{Koblitz1984,Robert2000}. For $a\in\Cp^\times$ the character $\chi_a(u):=e_p(au)$ is defined
on $B(a):=\{u\in\Cp:|au|_p<\rhop\}$ and satisfies
\begin{equation}\label{eq:charlip}
  |\chi_a(u)-\chi_a(v)|_p=|a|_p\,|u-v|_p \qquad (u,v\in B(a)).
\end{equation}

\subsection{The iterated integral}\label{sec:nested}

In \cite[Eq.~(13)]{HuKim2026} the integrand is $\chi_a$ of the total action, and the Dirac centre
for the variable $x_r$ depends on the next variable $x_{r+1}$. The following definition fixes the
order of evaluation.

\begin{definition}[Iterated Dirac integral]\label{def:nested}
Let $K=p^M\ge1$, let $S$ be a two-point action, let $f$ be a function on a subset of $\Cp$, and let
$x,y\in\Qp$. Put $\xi_0:=x$, $\xi_K:=y$, and for $r=K-1,K-2,\dots,1$,
\begin{equation}\label{eq:xi}
  \xi_r:=\frac{1}{r+1}\,x+\frac{r}{r+1}\,\xi_{r+1}.
\end{equation}
The iterated integral is defined by
\begin{equation}\label{eq:nested}
\begin{split}
  &\int\cdots\int f\Big(\sum_{i=0}^{K-1}S(x_i,x_{i+1})\Big)\prod_{r=1}^{K-1}\dd\delta_{d_r}(x_r)
   := f\Big(\sum_{i=0}^{K-1}S(\xi_i,\xi_{i+1})\Big),\\
  &\qquad d_r=\frac{1}{r+1}\,x+\frac{r}{r+1}\,x_{r+1},
\end{split}
\end{equation}
whenever the sum on the right lies in the domain of $f$.
\end{definition}

The variables are eliminated from $x_1$ outward to $x_{K-1}$, and the nodes are determined backward
from $\xi_K=y$; this is the only order compatible with the dependence of $d_r$ on $x_{r+1}$. The
substitution takes place inside $f$: the character of the total action is not split into a product
of slice characters, which in the presence of a potential would require stronger domain conditions.

\begin{lemma}[Collinear nodes]\label{lem:nested}
For $0\le r\le K$ one has $\xi_r=x+\frac rK(y-x)$. Hence \eqref{eq:nested} equals
$f\big(S^{(K)}_{\mathrm{eff}}(x,y)\big)$ with
\begin{equation}\label{eq:Seff_k}
  S^{(K)}_{\mathrm{eff}}(x,y)=\sum_{i=0}^{K-1}S\Big(x+\tfrac iK(y-x),\,x+\tfrac{i+1}{K}(y-x)\Big).
\end{equation}
\end{lemma}

\begin{proof}
By downward induction: if $\xi_{r+1}=x+\frac{r+1}{K}(y-x)$, then
$\xi_r=\frac{x}{r+1}+\frac{r}{r+1}\big[x+\frac{r+1}{K}(y-x)\big]=x+\frac rK(y-x)$.
\end{proof}

We write $S^{(n)}_{\mathrm{eff}}$ for \eqref{eq:Seff_k} at level $n$, that is, for $K=p^{N-n}$ and
the level-$n$ slice action.

\begin{definition}[Admissibility]\label{def:admissible}
The level-$n$ propagator
$U^{(n)}(y,t;x,t_0):=\chi_a\big(S^{(n)}_{\mathrm{eff}}(x,y)\big)$
is \emph{admissible} at $(x,y)$ if $|a\,S^{(n)}_{\mathrm{eff}}(x,y)|_p<\rhop$.
\end{definition}

\begin{proposition}[The free propagator of HK]\label{prop:HK_free}
Let $S(u,v)=\frac{m(v-u)^2}{2\epsilon_n}$ with $m\in\Qp^\times$. Then
$S^{(n)}_{\mathrm{eff}}(x,y)=\frac{m(y-x)^2}{2p^N}$ for every $n\le N$, and
$U^{(n)}(y,t;x,t_0)=e_p\big(\frac{am(y-x)^2}{2p^N}\big)$ whenever admissible. This is
\cite[Eq.~(17)]{HuKim2026}.
\end{proposition}

\begin{proof}
Each increment in \eqref{eq:Seff_k} equals $(y-x)/K$, so the sum is
$\frac{m}{2\epsilon_n}K\frac{(y-x)^2}{K^2}=\frac{m(y-x)^2}{2K\epsilon_n}$, and $K\epsilon_n=p^N$.
\end{proof}

\begin{convention}\label{conv:bare}
In Section~\ref{sec:perfect} the slice action is taken in the form used in HK,
\begin{equation}\label{eq:bare}
  S_n[i,i+1]=\frac{m}{2}\frac{(x_{i+1}-x_i)^2}{\epsilon_n}-V\Big(\frac{x_i+x_{i+1}}{2}\Big),
\end{equation}
in which $V$ carries no factor $\epsilon_n$. For $n=N$ this gives
$S^{(N)}_{\mathrm{eff}}(x,y)=\frac{m(y-x)^2}{2p^N}-V\big(\frac{x+y}{2}\big)$. The role of the
missing factor is examined in Proposition~\ref{prop:epsweight}.
\end{convention}

\section{Block decimation operators}\label{sec:RGop}

\subsection{Definitions}

\begin{definition}[Block decimation]\label{def:Kadanoff}
A \emph{node rule} is a family $\gamma=(\gamma_0,\dots,\gamma_p)$ of maps $\Qp^2\to\Qp$ (or
$\Cp^2\to\Cp$) with $\gamma_0(x,y)=x$ and $\gamma_p(x,y)=y$. For a two-point action $S$ put
\begin{equation}\label{eq:Kadanoff_def}
  \big(\sK^{(\gamma)}_pS\big)(x,y):=\sum_{r=0}^{p-1}S\big(\gamma_r(x,y),\gamma_{r+1}(x,y)\big).
\end{equation}
If $S$ is regarded as an action at scale $\epsilon$, then $\sK^{(\gamma)}_pS$ is regarded as an
action at scale $p\epsilon$.

The \emph{kinematic} rule is $\gamma^{\mathrm{free}}_r(x,y)=x+\frac rp(y-x)$, with operator
$\Kkin:=\sK^{(\gamma^{\mathrm{free}})}_p$. The \emph{variational} rule $\gamma^{S}$ consists of the
interior stationary points of the block sum,
\begin{equation}\label{eq:variational}
  \partial_{z_r}\Big[S(z_{r-1},z_r)+S(z_r,z_{r+1})\Big]_{z=\gamma^{S}(x,y)}=0,
  \qquad r=1,\dots,p-1,
\end{equation}
with $z_0=x$ and $z_p=y$, whenever this system has a unique solution in a specified domain
(Proposition~\ref{prop:tangent} and Definition~\ref{def:dyn}); we write
$\Kvar S:=\sK^{(\gamma^{S})}_pS$.
\end{definition}

For $S=\Sfree(\cdot,\cdot;\epsilon)$ the system \eqref{eq:variational} reads
$2z_r-z_{r-1}-z_{r+1}=0$, whose solution is $\gamma^{\mathrm{free}}$. The kinematic rule is thus
the variational rule of the free action, and by Lemma~\ref{lem:nested} it is the rule realized by
the Dirac centres of HK.

\begin{definition}[Perfect action]\label{def:perfect}
Let $\gamma$ be a node rule. A family $\{S^*_\epsilon\}_\epsilon$ of actions indexed by the scale
is \emph{scale-covariant}, or a \emph{perfect action for $\gamma$}, if
$\sK^{(\gamma)}_pS^*_\epsilon=S^*_{p\epsilon}$ for all $\epsilon$.
\end{definition}

\begin{lemma}[Scalar rescaling]\label{lem:Dp}
Let $m\in\Cp^\times$ and $\Sfree(x,y;\epsilon):=\frac{m(y-x)^2}{2\epsilon}$. Then
$\Kkin\Sfree(\cdot,\cdot;\epsilon)=\frac1p\Sfree(\cdot,\cdot;\epsilon)=\Sfree(\cdot,\cdot;p\epsilon)$.
Consequently, among the scalar maps $S\mapsto cS$ ($c\in\Cp$), the map
\begin{equation}\label{eq:Dp_def}
  \Dp S:=p\,S
\end{equation}
is the only one for which $\Rp:=\Dp\circ\Kkin$ fixes the free actions at fixed $\epsilon$.
\end{lemma}

\begin{proof}
$\Kkin\Sfree(\cdot,\cdot;\epsilon)=\sum_{r<p}\frac{m}{2\epsilon}\big(\frac{y-x}{p}\big)^2
=\frac1p\Sfree(\cdot,\cdot;\epsilon)$.
\end{proof}

\begin{remark}[Other rescalings; the running character]\label{rem:rescaling}
(i) A rescaling may also involve a dilation of the coordinates, $S\mapsto c\,S(\lambda\,\cdot,
\lambda\,\cdot)$ with $\lambda\in\Qp^\times$; it fixes the free actions when $c\lambda^2=p$. We do
not use such maps. For $|\lambda|_p\ne1$ they move the support $\Zp$ of the potentials of
Section~\ref{sec:perfect}, and for $|\lambda|_p=1$ they differ from $\Dp$ by a dilation that
commutes with $\Kkin$. A dilation by $\sqrt p$ with $c=1$ is excluded because
$\sqrt p\notin\Qp$.

(ii) Since $\chi_a(pS)=\chi_{pa}(S)$, applying $\Dp$ at fixed $a$ is the same as keeping the action
and replacing $a$ by $pa$. The parameter $a$ plays the role of $1/\hbar$, so the rescaled flow at
fixed $a$ is the unrescaled flow with a character parameter that changes with the scale.
\end{remark}

\begin{remark}[Scale-covariant families and fixed points]\label{rem:two_notions}
For $m\ne0$ the free action at fixed $\epsilon$ is a fixed point of $\Rp$, while the family
$\epsilon\mapsto\Sfree(\cdot,\cdot;\epsilon)$ is a perfect action for $\gamma^{\mathrm{free}}$. The
two notions differ: a constant family $S_\epsilon=S$ is scale-covariant if and only if $\Kkin S=S$,
and then $\Rp S=pS$, so $S$ is a fixed point of $\Rp$ only if $S=0$. In
Section~\ref{sec:perfect} perfectness is formulated as scale covariance. The covariance identity
alone does not determine the limit action ($W\equiv0$ also satisfies it); as in
\cite{Hasenfratz1994}, the limit is selected by the flow that starts from the bare action
\eqref{eq:bare}, and Proposition~\ref{prop:char} expresses this selection as a boundary condition
at large separation.
\end{remark}

\subsection{Continuity and unboundedness}\label{sec:spaces}

\begin{proposition}[Continuity; shift of the polydisc]\label{prop:continuous}
Let $\sE:=C(\Qp^2;\Cp)$ with the topology of uniform convergence on compact sets, a
non-Archimedean Fr\'echet space. For $\nu\in\Z$ let
$\sA_\nu:=\Cp\langle u/p^\nu,v/p^\nu\rangle$ be the Tate algebra of power series
$\sum_{i,j\ge0}b_{ij}u^iv^j$ that converge on the polydisc
\[
  D_{p^{-\nu}}^2=\{(u,v)\in\Cp^2:|u|_p,|v|_p\le p^{-\nu}\},
\]
with the Gauss norm
$\|S\|_\nu:=\sup_{D_{p^{-\nu}}^2}|S|_p=\max_{i,j}|b_{ij}|_p\,p^{-\nu(i+j)}$.
\begin{enumerate}
\item For $0\le r\le p$,
  $|\gamma^{\mathrm{free}}_r(u,v)|_p\le\max\{|u|_p,\,p\,|v-u|_p\}$. Hence
  $\gamma^{\mathrm{free}}_r$ maps $D_R^2$ into $D_{pR}$ for every $R>0$, and no polydisc $D_R^2$
  is mapped into itself by all $\gamma^{\mathrm{free}}_r$.
\item $\Kkin$ is a continuous linear endomorphism of $\sE$, and it maps $\sA_\nu$ into
  $\sA_{\nu+1}$ with $\|\Kkin S\|_{\nu+1}\le\|S\|_\nu$.
\item $\Kkin$ is not bounded by $1$ on any $\sA_\nu$-norm restricted to polynomials: for
  $S=(v-u)^2$ one has $\Kkin S=p^{-1}S$ and $\|\Kkin S\|_\nu=p\,\|S\|_\nu$.
\end{enumerate}
\end{proposition}

\begin{proof}
(1) Since $|r/p|_p\le p$, the ultrametric inequality gives the bound. Moreover
$\gamma^{\mathrm{free}}_1(0,v)=v/p$ has norm $p|v|_p$.
(2) The maps $\gamma^{\mathrm{free}}_r$ are affine, hence continuous, and they map compact sets to
compact sets; this gives continuity on $\sE$. If $(u,v)\in D_{p^{-(\nu+1)}}^2$, then by (1) all
nodes lie in $D_{p^{-\nu}}$. For a polynomial $P$ this gives
$\|\Kkin P\|_{\nu+1}\le\max_{r}\sup_{D_{p^{-\nu}}^2}|P|_p=\|P\|_\nu$, and polynomials are dense in
$\sA_\nu$.
(3) $|p^{-1}|_p=p$.
\end{proof}

\begin{proposition}[The rescaled map is not a contraction]\label{prop:nogo}
Let $\sX$ be a $\Cp$-vector space of functions on $\Qp^2$ that is invariant under $\Rp$, and let
$\|\cdot\|$ be a norm on $\sX$ with $\|cS\|=|c|_p\|S\|$. If $(v-u)^b\in\sX$ for an integer
$b\ge2$, then every bound for $\Rp$ is at least $p^{\,b-2}$. In particular $\|\Rp\|\ge1$ whenever
$\sX$ contains a free action, and $\Rp$ is unbounded whenever $\sX$ contains $(v-u)^b$ for
infinitely many $b$.
\end{proposition}

\begin{proof}
$\Rp\big[(v-u)^b\big]=p\sum_{r<p}\big(\frac{v-u}{p}\big)^b=p^{2-b}(v-u)^b$, and
$|p^{2-b}|_p=p^{\,b-2}$.
\end{proof}

\begin{remark}\label{rem:nogo}
The free actions form a line of fixed points of $\Rp$ parametrized by the mass, and the directions
$(v-u)^b$ with $b\ge4$ are expanded. The convergence statements below therefore do not follow from
Banach's fixed-point theorem applied to $\Rp$. They are proved for the interaction part of
potentials supported on $\Zp$, in the supremum norm on $\Zp^2$ (Section~\ref{sec:perfect}), and
for quadratic actions (Section~\ref{sec:oscillator}).
\end{remark}

\subsection{Linearization of the variational decimation}\label{sec:tangent}

\begin{proposition}[Linearization]\label{prop:tangent}
Let $m\in\Cp^\times$, $\epsilon\in\Q^\times$, $\nu\in\Z$, and $\Lambda:=|m/\epsilon|_p$. Let
$U\in\sA_\nu$ satisfy
\begin{equation}\label{eq:small}
  \|U\|_\nu<\Lambda\,p^{-1-2\nu}.
\end{equation}
\begin{enumerate}
\item For every $(x,y)\in D^2_{p^{-(\nu+1)}}$ the system \eqref{eq:variational} for
  $S=\Sfree(\cdot,\cdot;\epsilon)-U$ has exactly one solution $\gamma^{U}(x,y)$ in
  $D_{p^{-\nu}}^{\,p-1}$, and
  \[
    \max_{1\le r\le p-1}\big|\gamma^{U}_r(x,y)-\gamma^{\mathrm{free}}_r(x,y)\big|_p
    \le\Lambda^{-1}p^{\nu+1}\|U\|_\nu .
  \]
\item $\Kvar[\Sfree(\cdot,\cdot;\epsilon)-U]$ belongs to $\sA_{\nu+1}$, and
  \begin{equation}\label{eq:tangent}
    \big\|\Kvar[\Sfree(\cdot,\cdot;\epsilon)-U]-\Sfree(\cdot,\cdot;p\epsilon)+\Kkin U\big\|_{\nu+1}
    \le\Lambda^{-1}p^{2\nu+2}\|U\|_\nu^2 .
  \end{equation}
\end{enumerate}
In particular $\Kvar$, as a map from the ball \eqref{eq:small} around
$\Sfree(\cdot,\cdot;\epsilon)$ in $\sA_\nu$ to $\sA_{\nu+1}$, is Fr\'echet differentiable at the
free action with derivative $\Kkin$, and the derivative of $\Dp\circ\Kvar$ there is
$\sT:=p\,\Kkin$.
\end{proposition}

\begin{proof}
We use two elementary estimates for $f=\sum b_{ij}u^iv^j\in\sA_\nu$, both verified monomial by
monomial: $\|\partial_uf\|_\nu,\|\partial_vf\|_\nu\le p^{\nu}\|f\|_\nu$, and
$|f(w)-f(w')|_p\le p^{\nu}\|f\|_\nu\,|w-w'|_p$ for $w,w'\in D^2_{p^{-\nu}}$ (maximum norm on
$\Cp^2$).

Fix $(x,y)\in D^2_{p^{-(\nu+1)}}$ and put $z_0=x$, $z_p=y$. The stationarity equations for
$\Phi(z):=\sum_{r=0}^{p-1}\big[\Sfree(z_r,z_{r+1};\epsilon)-U(z_r,z_{r+1})\big]$ are
\[
\begin{aligned}
  \frac m\epsilon\,(2z_r-z_{r-1}-z_{r+1})&=G_r(z),\\
  G_r(z):=\partial_2U(z_{r-1},z_r)+\partial_1U(z_r,z_{r+1}),&\qquad 1\le r\le p-1 .
\end{aligned}
\]
Write $z_r=\gamma^{\mathrm{free}}_r(x,y)+\zeta_r$ with $\zeta_0=\zeta_p=0$. Since
$\gamma^{\mathrm{free}}$ solves the homogeneous equations, the system is equivalent to
\[
  \zeta=\Psi(\zeta):=\frac\epsilon m\,\Delta_{p-1}^{-1}\,G\big(\gamma^{\mathrm{free}}+\zeta\big),
\]
where $\Delta_{p-1}$ is the $(p-1)\times(p-1)$ matrix with $2$ on the diagonal and $-1$ next to
it. Its inverse has entries $\min(i,j)\,(p-\max(i,j))/p$, all of norm $p$.

Let $B:=\{\zeta\in\Cp^{\,p-1}:\max_r|\zeta_r|_p\le p^{-\nu}\}$. By
Proposition~\ref{prop:continuous}(1) the free nodes lie in $D_{p^{-\nu}}$, so for $\zeta\in B$ all
arguments of $G$ lie in $D^2_{p^{-\nu}}$ and $|G_r|_p\le p^\nu\|U\|_\nu$. Hence
$|\Psi(\zeta)|_p\le\Lambda^{-1}p^{\nu+1}\|U\|_\nu<p^{-\nu}$ by \eqref{eq:small}, so $\Psi$ maps
$B$ into itself. Since $\partial_iU\in\sA_\nu$ has norm at most $p^\nu\|U\|_\nu$,
\[
  |\Psi(\zeta)-\Psi(\zeta')|_p\le\Lambda^{-1}p^{2\nu+1}\|U\|_\nu\,|\zeta-\zeta'|_p ,
\]
and the constant is $<1$ by \eqref{eq:small}. Thus $\Psi$ has a unique fixed point in the complete
set $B$. Every solution of the system in $D_{p^{-\nu}}^{\,p-1}$ corresponds to some $\zeta\in B$,
by the ultrametric inequality; this proves (1), the bound being $|\zeta|_p=|\Psi(\zeta)|_p$.

The iterates $\zeta^{(0)}=0$, $\zeta^{(j+1)}=\Psi(\zeta^{(j)})$, regarded as functions of $(x,y)$,
have components in $\sA_{\nu+1}$ of norm at most $p^{-\nu}$, because $\sA_{\nu+1}$ is closed under
composition with elements of $\sA_\nu$ whose arguments have norm at most $p^{-\nu}$. The estimates
above hold uniformly on $D^2_{p^{-(\nu+1)}}$, that is, in the norm $\|\cdot\|_{\nu+1}$, so the
iterates form a Cauchy sequence in the Banach space $\sA_{\nu+1}$. The fixed point therefore has
components in $\sA_{\nu+1}$, and so does $\Kvar[\Sfree-U]=\Phi(\gamma^{\mathrm{free}}+\zeta)$.

For (2), the free part of $\Phi$ is quadratic, and the term linear in $\zeta$ vanishes because
$\sum_r(\zeta_{r+1}-\zeta_r)=0$:
\[
  \sum_{r=0}^{p-1}\Sfree(z_r,z_{r+1};\epsilon)
  =\Sfree(x,y;p\epsilon)+\frac{m}{2\epsilon}\sum_{r=0}^{p-1}(\zeta_{r+1}-\zeta_r)^2 .
\]
Therefore $\Kvar[\Sfree-U]=\Sfree(\cdot,\cdot;p\epsilon)-\Kkin U+R$ with
\[
  R=\frac{m}{2\epsilon}\sum_{r=0}^{p-1}(\zeta_{r+1}-\zeta_r)^2
  -\sum_{r=0}^{p-1}\Big[U\big(\gamma^{\mathrm{free}}_r+\zeta_r,\gamma^{\mathrm{free}}_{r+1}
  +\zeta_{r+1}\big)-U\big(\gamma^{\mathrm{free}}_r,\gamma^{\mathrm{free}}_{r+1}\big)\Big].
\]
The first sum is bounded by $\Lambda|\zeta|_p^2\le\Lambda^{-1}p^{2\nu+2}\|U\|_\nu^2$ and the
second by $p^\nu\|U\|_\nu|\zeta|_p\le\Lambda^{-1}p^{2\nu+1}\|U\|_\nu^2$.
\end{proof}

\begin{remark}[Range of the linear regime]\label{rem:window}
For $\epsilon=\epsilon_n$ one has $\Lambda=|m|_p\,p^{n}$, so \eqref{eq:small} reads
$\|U\|_\nu<|m|_p\,p^{\,n-1-2\nu}$. The admissible perturbations shrink to zero under refinement
($n\to-\infty$) and grow under coarsening. The potentials of Section~\ref{sec:perfect} are cut off
outside $\Zp$ and do not belong to any $\sA_\nu$, so Proposition~\ref{prop:tangent} does not apply
to them; there $\Kkin$ is studied as the HK prescription in its own right.
\end{remark}

\begin{definition}[Relevance]\label{def:relevance}
Let $\lambda\in\Cp$ be an eigenvalue of the linearization of a rescaled decimation map at a fixed
point. The corresponding eigenvector is called \emph{$p$-adically relevant}, \emph{marginal}, or
\emph{irrelevant} if $|\lambda|_p>1$, $=1$, or $<1$. When $\lambda\in\Q$ we use the same terms
with $|\lambda|_\infty$ for the real norm.
\end{definition}

\subsection{Iteration and the free line}

\begin{lemma}[Iteration of the kinematic decimation]\label{lem:semigroup}
For $M\ge0$ and $K=p^M$ one has $(\Kkin)^{M}S=S^{(K)}_{\mathrm{eff}}$ in the sense of
\eqref{eq:Seff_k}. Thus $M$ kinematic steps equal the $(p^M-1)$-fold iterated Dirac integral of
Definition~\ref{def:nested}.
\end{lemma}

\begin{proof}
By induction on $M$. Applying $\Kkin$ to $(\Kkin)^MS$ divides each interval $[z_j,z_{j+1}]$,
$z_j=x+\frac jp(y-x)$, into $p^M$ equal parts with nodes
$x+\frac{jp^M+i}{p^{M+1}}(y-x)$, $0\le i\le p^M$. As $(j,i)$ runs over
$\{0,\dots,p-1\}\times\{0,\dots,p^M-1\}$, the numerator $jp^M+i$ runs once over
$\{0,\dots,p^{M+1}-1\}$.
\end{proof}

\begin{proposition}[The free line]\label{prop:free_FP}
For every $m\in\Cp^\times$ and $\epsilon\in\Q^\times$,
$\Kkin\Sfree(\cdot,\cdot;\epsilon)=\Kvar\Sfree(\cdot,\cdot;\epsilon)=\Sfree(\cdot,\cdot;p\epsilon)$
and $\Rp\Sfree(\cdot,\cdot;\epsilon)=\Sfree(\cdot,\cdot;\epsilon)$. Consequently, for $V\equiv0$,
$S^{(n)}_{\mathrm{eff}}(x,y)=\frac{m(y-x)^2}{2p^N}$ for all $n\le N$.
\end{proposition}

\begin{proof}
Lemma~\ref{lem:Dp}, Lemma~\ref{lem:semigroup}, and the equality
$\gamma^{\Sfree}=\gamma^{\mathrm{free}}$.
\end{proof}

\begin{remark}\label{rem:classical}
Because \eqref{eq:dirac} evaluates at one node per slice, the decimation generates no multi-point
couplings and needs no truncation. Lemma~\ref{lem:semigroup} concerns the fixed rule
$\gamma^{\mathrm{free}}$. Under $\Kvar$ the rule changes with the action at each step; for
quadratic actions the composition is described in Proposition~\ref{prop:path}.
\end{remark}

\section{Perfect actions for potentials supported on $\Zp$}\label{sec:perfect}

By Lemma~\ref{lem:semigroup} and Proposition~\ref{prop:free_FP}, the level-$n$ effective action
for the slice action \eqref{eq:bare} is, with $M=N-n\ge0$,
\begin{equation}\label{eq:Weff}
\begin{aligned}
  S^{(n)}_{\mathrm{eff}}(x,y)&=\frac{m(y-x)^2}{2p^N}-W_M(x,y),\\
  W_M(x,y)&:=\sum_{i=0}^{p^M-1}V\Big(x+\frac{2i+1}{2p^M}(y-x)\Big).
\end{aligned}
\end{equation}
We also write $W^{(n)}:=W_{N-n}$. The substitution $i\mapsto p^M-1-i$ shows that
$W_M(y,x)=W_M(x,y)$.

\medskip
\noindent\textbf{Hypothesis (H$_0$).}
$V:\Qp\to\Cp$ vanishes outside $\Zp$.

\smallskip
\noindent\textbf{Hypothesis (H).}
(H$_0$) holds, and on $\Zp$ the function $V$ is given by a series $V(z)=\sum_{j\ge0}c_jz^j$ with
$c_j\to0$. We write
\[
  \nT{V}:=\sup_{j\ge0}|c_j|_p,\qquad \tau_k:=\sup_{j\ge k}|c_j|_p\quad(k\ge0),
\]
so that $\tau_k$ decreases to $0$. Since $\Zp$ is open and closed in $\Qp$, $V$ is continuous on
$\Qp$ and $\sup_{\Qp}|V|_p\le\nT{V}$.

\begin{remark}[Function spaces]\label{rem:interaction}
Under (H) each $W_M$ is a bounded continuous function on $\Qp^2$ with $\sup|W_M|_p\le\nT{V}$, and
$\Kkin$ does not increase the supremum norm on bounded functions on $\Qp^2$. The operator $\Kkin$
does not act on $C(\Zp^2;\Cp)$, because the nodes of a pair $(x,y)\in\Zp^2$ with $|y-x|_p=1$
leave $\Zp$. The recursion is therefore carried out on $\Qp^2$, and the uniform estimates are
stated for the restrictions to $\Zp^2$.
\end{remark}

\begin{example}[The support hypothesis is needed]\label{ex:nosupp}
Let $V(z)=z^2$ on all of $\Qp$. From $\sum_{i<K}(2i+1)=K^2$ and
$\sum_{i<K}(2i+1)^2=K(4K^2-1)/3$ with $K=p^M$,
\[
  W_M(x,y)=p^Mx^2+p^Mx(y-x)+\frac{(4p^{2M}-1)(y-x)^2}{12\,p^M}.
\]
For $x\ne y$ the last term has norm $|y-x|_p^2\,p^M\,|1/12|_p\to\infty$. In terms of
Section~\ref{sec:spectrum}: writing $q=\frac{x+y}2$ and $\eta=y-x$, one has
$q^2=\big(q^2+\frac1{12}\eta^2\big)-\frac1{12}\eta^2$, a sum of eigenvectors of $\Kkin$ with
eigenvalues $p$ and $p^{-1}$, so that
$W_M=p^M\big(q^2+\frac1{12}\eta^2\big)-\frac{1}{12\,p^M}\eta^2$. The divergence comes from the
component along $\eta^2$.
\end{example}

\subsection{Arithmetic estimates}

Let $B_j$ denote the Bernoulli numbers, with $B_1=-\frac12$.

\begin{lemma}[Power sums]\label{lem:Faulhaber}
For all integers $L\ge1$ and $k\ge0$,
\begin{equation}\label{eq:volk_claim}
  v_p\Big(p^{-L}\sum_{i=0}^{p^L-1}i^k-B_k\Big)\ \ge\ L-1 .
\end{equation}
Consequently $\big|\sum_{i=0}^{p^L-1}i^k\big|_p\le p^{1-L}$ for $k\ge1$, while for $k=0$ the sum
equals $p^L$.
\end{lemma}

\begin{proof}
For $k=0$ the left side of \eqref{eq:volk_claim} vanishes. For $k\ge1$, Faulhaber's formula
$\sum_{i=0}^{n-1}i^k=\sum_{j=0}^{k}\frac1{j+1}\binom kjB_{k-j}\,n^{j+1}$ gives
\[
  p^{-L}\sum_{i=0}^{p^L-1}i^k-B_k=\sum_{j=1}^{k}\frac{1}{j+1}\binom{k}{j}B_{k-j}\,p^{Lj}.
\]
By the Clausen--von Staudt theorem $pB_i\in\Zp$ for all $i$ \cite{Koblitz1984,Robert2000}, so the
$j$-th term has valuation at least $Lj-e-1$ with $e:=v_p(j+1)$. If $e=0$ this is $\ge L-1$. If
$e\ge1$, then $j+1\ge p^e\ge3^e\ge e+2$, so $j\ge e+1\ge2$ and
$Lj-e-1=(L-1)j+(j-e-1)\ge2(L-1)\ge L-1$. For the second statement write the sum as
$p^L(B_k+\rho)$ with $v_p(\rho)\ge L-1$ and use $v_p(B_k)\ge-1$.
\end{proof}

\begin{lemma}[Affine substitution]\label{lem:affine}
Assume (H) and let $\alpha,\delta\in\Zp$. Then $h(z):=V(\alpha+\delta z)$, $z\in\Zp$, has an
expansion $h(z)=\sum_{k\ge0}b_kz^k$ with
\[
\begin{aligned}
  b_k&=\delta^k\sum_{j\ge k}c_j\binom jk\alpha^{j-k},\\
  |b_k|_p&\le|\delta|_p^k\,\tau_k\le\nT{V}.
\end{aligned}
\]
\end{lemma}

\begin{proof}
In the double series $\sum_{j}\sum_{k\le j}c_j\binom jk\alpha^{j-k}\delta^kz^k$ the terms are
bounded by $|c_j|_p\to0$, which justifies the rearrangement. The bound follows from
$|\binom jk|_p\le1$, $|\alpha|_p\le1$ and $|c_j|_p\le\tau_k$ for $j\ge k$.
\end{proof}

\begin{corollary}\label{cor:scaled_Faulhaber}
In the setting of Lemma~\ref{lem:affine}, for $M\ge1$,
\begin{equation}\label{eq:cor_bound}
  \Big|\sum_{i=0}^{p^M-1}h(i)\Big|_p\le p^{1-M}\,\nT{V}.
\end{equation}
If moreover $|\delta|_p\le p^{-1}$, the bound improves to $p^{-M}\nT{V}$.
\end{corollary}

\begin{proof}
Write the sum as $\sum_{k\ge0}b_kP_k$ with $P_k:=\sum_{i<p^M}i^k$. The term $k=0$ is $p^Mb_0$, of
norm $\le p^{-M}\nT{V}$. For $k\ge1$, Lemma~\ref{lem:Faulhaber} gives
$|b_kP_k|_p\le|\delta|_p^k\,p^{1-M}\nT{V}$, which is $\le p^{1-M}\nT{V}$ in general and
$\le p^{-M}\nT{V}$ when $|\delta|_p\le p^{-1}$.
\end{proof}

\begin{remark}\label{rem:sharpening}
Lemma~\ref{lem:Faulhaber} alone gives $p^{1-M}$, and this cannot be improved
(Example~\ref{ex:A_Faulhaber}). The gain to $p^{-M}$ comes from the factor $\delta^k$ in the
Taylor coefficients of $V(\alpha+\delta z)$. Near the diagonal, $\delta=(y-x)/p^M$ satisfies
$|\delta|_p\le p^{-1}$ as long as $M<v_p(y-x)$.
\end{remark}

\subsection{Stabilization, characterization, and uniform convergence}

\begin{proposition}[Recursion and stabilization]\label{prop:FP}
Assume (H$_0$) and put $z_j:=x+\frac jp(y-x)$ for $0\le j\le p$.
\begin{enumerate}
\item $W_{M+1}(x,y)=\sum_{j=0}^{p-1}W_M(z_j,z_{j+1})$; equivalently
  $\Kkin\big(\Sfree(\cdot,\cdot;\epsilon)-W_M\big)=\Sfree(\cdot,\cdot;p\epsilon)-W_{M+1}$.
\item If $x\ne y$ and $p^{M+1}|y-x|_p>\max\{1,|x|_p\}$, then $W_{M'}(x,y)=W_M(x,y)$ for all
  $M'\ge M$. If $x=y$, then $W_M(x,x)=p^MV(x)\to0$. Hence
  $W^*(x,y):=\lim_{M\to\infty}W_M(x,y)$ exists for all $(x,y)\in\Qp^2$.
\item For $x,y\in\Zp$ with $x\ne y$ and $s:=v_p(y-x)$,
  \begin{equation}\label{eq:Wstar}
    W^*(x,y)=W_s(x,y)=\sum_{i=0}^{p^s-1}V\Big(x+\frac{2i+1}{2p^s}(y-x)\Big),
    \qquad W^*(x,x)=0 .
  \end{equation}
\item $\Kkin W^*=W^*$ on $\Qp^2$. Hence the family
  $S^*_\epsilon(u,v):=\frac{m(v-u)^2}{2\epsilon}-W^*(u,v)$ is a perfect action for
  $\gamma^{\mathrm{free}}$: $\Kkin S^*_\epsilon=S^*_{p\epsilon}$.
\end{enumerate}
\end{proposition}

\begin{proof}
(1) With $i'=jp^M+i$ one has
$x+\frac{2i'+1}{2p^{M+1}}(y-x)=z_j+\frac{2i+1}{2p^M}(z_{j+1}-z_j)$; sum over $(j,i)$ and use
Proposition~\ref{prop:free_FP}.

(2) Let $F_M:=\{\frac{2i+1}{2p^M}:0\le i<p^M\}$. Since $p$ is odd,
$\frac{2i+1}{2p^M}=\frac{p(2i+1)}{2p^{M+1}}$ lies in $F_{M+1}$, so $F_M\subset F_{M+1}$, and the
elements $\lambda\in F_{M+1}\setminus F_M$ are those with $p\nmid(2i'+1)$, for which
$|\lambda|_p=p^{M+1}$. For such $\lambda$,
$|\lambda(y-x)|_p=p^{M+1}|y-x|_p>\max\{1,|x|_p\}$, hence
$|x+\lambda(y-x)|_p=|\lambda(y-x)|_p>1$ and $V(x+\lambda(y-x))=0$. Thus
$W_{M+1}(x,y)=W_M(x,y)$, and the hypothesis persists when $M$ increases. The diagonal case is
clear.

(3) For $x,y\in\Zp$ the condition in (2) reads $p^{M+1-s}>1$, that is, $M\ge s$.

(4) Let $M\to\infty$ in the finite sum of (1), using (2) for each term, and apply
Proposition~\ref{prop:free_FP} to the kinetic part.
\end{proof}

\begin{proposition}[Characterization of $W^*$]\label{prop:char}
Assume (H$_0$). The function $W^*$ of Proposition~\ref{prop:FP} is the unique function
$W:\Qp^2\to\Cp$ with the following two properties:
\begin{enumerate}
\item[(a)] $\Kkin W=W$;
\item[(b)] $W(x,y)=V\big(\frac{x+y}2\big)$ whenever $x\ne y$ and
  $|y-x|_p\ge\max\{1,|x|_p\}$.
\end{enumerate}
\end{proposition}

\begin{proof}
$W^*$ satisfies (a) by Proposition~\ref{prop:FP}(4). For $x,y\in\Qp$ the inequality in (b) is
equivalent to $p\,|y-x|_p>\max\{1,|x|_p\}$, and Proposition~\ref{prop:FP}(2) with $M=0$ then gives
$W^*(x,y)=W_0(x,y)=V\big(\frac{x+y}2\big)$.

Conversely, let $W$ satisfy (a) and (b). On the diagonal, (a) reads $W(x,x)=pW(x,x)$, so
$W(x,x)=0=W^*(x,x)$. Let $x\ne y$ and choose $M\ge0$ with $p^M|y-x|_p\ge\max\{1,|x|_p\}$. By (a)
and Lemma~\ref{lem:semigroup},
$W(x,y)=\sum_{i=0}^{p^M-1}W(\xi_i,\xi_{i+1})$ with $\xi_i=x+ip^{-M}(y-x)$. Here
$|\xi_{i+1}-\xi_i|_p=p^M|y-x|_p$ and
$|\xi_i|_p\le\max\{|x|_p,p^M|y-x|_p\}=p^M|y-x|_p$, so (b) applies to each pair
$(\xi_i,\xi_{i+1})$ and
$W(x,y)=\sum_iV\big(\frac{\xi_i+\xi_{i+1}}2\big)=W_M(x,y)$. Since
$p^{M+1}|y-x|_p>\max\{1,|x|_p\}$, Proposition~\ref{prop:FP}(2) gives $W_M(x,y)=W^*(x,y)$.
\end{proof}

\begin{theorem}[Stabilization and uniform convergence]\label{thm:uniform_RG}
Assume (H), and let $n\le N$, $M=N-n$.
\begin{enumerate}
\item Let $x\ne y$ in $\Zp$ and $s=v_p(y-x)$. Then $W^{(n)}(x,y)=W^*(x,y)$ for all
  $n\le N-s$; in particular, for $s=0$ one has $W^{(n)}(x,y)=V(\frac{x+y}2)$ for all
  $n\le N$. For $N-s<n\le N$ one has $|W^{(n)}(x,y)|_p\le p^{-(N-n)}\nT{V}$.
\item $W^*$ is continuous and symmetric on $\Zp^2$,
  $|W^*(x,y)|_p\le\min\{1,p\,|y-x|_p\}\,\nT{V}$, and
  \begin{equation}\label{eq:rate}
    \sup_{x,y\in\Zp}\big|W^{(n)}(x,y)-W^*(x,y)\big|_p\ \le\ p^{-(N-n)}\,\nT{V}
    \qquad(n\le N).
  \end{equation}
  Neither the exponent nor the constant can be improved (Remark~\ref{rem:sharp}).
\item Assume $|a|_p\max\{|m|_p\,p^N,\nT{V}\}<\rhop$. Then
  $U^{(n)}=\chi_a(S^{(n)}_{\mathrm{eff}})$ and $U^*:=\chi_a(S^*_{p^N})$ are admissible on
  $\Zp^2$, and
  \begin{equation}\label{eq:proprate}
    \sup_{x,y\in\Zp}\big|U^{(n)}(y,t;x,t_0)-U^*(y,t;x,t_0)\big|_p\le|a|_p\,p^{-(N-n)}\,\nT{V}.
  \end{equation}
\end{enumerate}
\end{theorem}

\begin{proof}
(1) The first statement is Proposition~\ref{prop:FP}(3). For $0\le M<s$ put
$\delta:=(y-x)/p^M$, so $|\delta|_p=p^{M-s}\le p^{-1}$. The midpoints are
$\alpha+\delta i$ with $\alpha=x+\frac\delta2\in\Zp$, and Corollary~\ref{cor:scaled_Faulhaber}
gives $|W_M(x,y)|_p\le p^{-M}\nT{V}$ for $M\ge1$; for $M=0$ the bound is trivial.

(2) If $s\ge1$, put $\varrho:=p^{-s}(y-x)\in\Zp^\times$; then
$W^*(x,y)=\sum_{i<p^s}V(x+\frac\varrho2+\varrho i)$, and \eqref{eq:cor_bound} with $M=s$
gives $|W^*(x,y)|_p\le p^{1-s}\nT{V}=p\,|y-x|_p\nT{V}$. The bound $|W^*|_p\le\nT{V}$ is
clear. For \eqref{eq:rate} with $M=0$, use $|W_0-W^*|_p\le\max\{|W_0|_p,|W^*|_p\}$. Let $M\ge1$.
If $s\le M$ the difference vanishes by (1). If $s>M$ (including $x=y$), then
$|W_M(x,y)|_p\le p^{-M}\nT{V}$ by (1), or because $W_M(x,x)=p^MV(x)$, and
$|W^*(x,y)|_p\le p^{1-s}\nT{V}\le p^{-M}\nT{V}$, or $W^*(x,x)=0$. Continuity of $W^*$ follows
from uniform convergence, and symmetry from that of $W_M$.

(3) On $\Zp^2$ the kinetic term has norm $\le|m|_p\,p^N$ and $|W^{(n)}|_p,|W^*|_p\le\nT{V}$, so
all actions lie in $B(a)$. Apply \eqref{eq:charlip} and \eqref{eq:rate}.
\end{proof}

\begin{remark}[Sharpness; $W^*$ is not analytic]\label{rem:sharp}
Let $V=\mathbf 1_{\Zp}$, so $\nT{V}=1$. Then $W^{(n)}(x,y)=p^{\min\{s,N-n\}}$ on $\Zp^2$,
$W^*(x,y)=p^s$ for $x\ne y$, and $W^*(x,x)=0$. For $s>M$ the difference
$p^M(1-p^{s-M})$ has norm exactly $p^{-M}$, so \eqref{eq:rate} is an equality. In this example
$W^*(x,y)$ is the rational number $|y-x|_p^{-1}$ off the diagonal; it is continuous on $\Zp^2$
but locally constant and non-constant off the diagonal, hence not given by a single power series.
This is why the limit is taken in the supremum norm and not in a Tate algebra. For non-constant
$V$ the bound can also be attained off the diagonal at the last step before stabilization,
$M=s-1$: this happens when $|W_{s-1}(x,y)|_p<|W^*(x,y)|_p=p^{1-s}\nT{V}$, as in
Example~\ref{ex:A_rate}. Saturation of the bound for $W^*$ alone does not suffice; for $p=3$,
$V=(1-z^2)\mathbf 1_{\Z_3}$ and $(x,y)=(0,9)$ one finds $W_1=-\frac{303}4$ and
$W^*=-\frac{933}4$, both of norm $3^{-1}$, with difference of norm $3^{-2}$.
\end{remark}

\begin{remark}[Multiplicativity along the chain]\label{rem:chapman}
By Proposition~\ref{prop:FP}(4), with $z_j=x+\frac jp(y-x)$ and whenever each factor is
admissible, we have
$\chi_a\big(S^*_{p^N}(x,y)\big)=\prod_{j=0}^{p-1}\chi_a\big(S^*_{p^{N-1}}(z_j,z_{j+1})\big)$.
\end{remark}

\begin{remark}\label{rem:Ndep}
The hypothesis of Theorem~\ref{thm:uniform_RG}(3) contains $p^N$ and fails for large $N$ at fixed
$a$. The limit $n\to-\infty$ is taken at fixed $N$.
\end{remark}

\subsection{Volkenborn asymptotics near the diagonal}

The Volkenborn integral of a strictly differentiable function $f:\Zp\to\Cp$ is
$\int_{\Zp}f\dd\mu_{\mathrm{Vol}}:=\lim_{L\to\infty}p^{-L}\sum_{i=0}^{p^L-1}f(i)$, and
$\int_{\Zp}z^k\dd\mu_{\mathrm{Vol}}=B_k$ \cite{Schikhof1984,Robert2000}.

\begin{proposition}[Volkenborn form of $W^*$]\label{prop:Volkenborn}
Assume (H). Let $x\ne y$ in $\Zp$ with $s=v_p(y-x)\ge1$, put
$\varrho:=p^{-s}(y-x)\in\Zp^\times$ and $g_{x,\varrho}(z):=V(x+\frac\varrho2+\varrho z)$. Then
\begin{equation}\label{eq:volk}
  \Big|W^*(x,y)-p^s\!\int_{\Zp}g_{x,\varrho}\dd\mu_{\mathrm{Vol}}\Big|_p
  \ \le\ p^{1-2s}\,\nT{V}.
\end{equation}
\end{proposition}

\begin{proof}
By Lemma~\ref{lem:affine}, $g_{x,\varrho}(z)=\sum_kb_kz^k$ with $|b_k|_p\le\tau_k$. Since
$|B_k|_p\le p$, the series $\sum_kb_kB_k$ converges, and by the uniform estimate
\eqref{eq:volk_claim} the limit $L\to\infty$ may be taken term by term, so
$\int_{\Zp}g_{x,\varrho}\dd\mu_{\mathrm{Vol}}=\sum_kb_kB_k$. By \eqref{eq:Wstar} and
\eqref{eq:volk_claim} with $L=s$,
\[
  W^*(x,y)-p^s\sum_kb_kB_k
  =p^s\sum_{k}b_k\Big(p^{-s}\sum_{i<p^s}i^k-B_k\Big),
\]
which has norm at most $p^{-s}\cdot p^{1-s}\nT{V}$.
\end{proof}

The factor $p^s=|y-x|_p^{-1}$ is the number of midpoints that remain in $\Zp$, and the
Volkenborn integral is the limiting average of $V$ over them.

\subsection{The weight of the potential}

A Riemann-sum discretization of $\int L\dd t$ would weight the potential by $\epsilon_n$:
$S^{\epsilon}_n[i,i+1]:=\frac{m(x_{i+1}-x_i)^2}{2\epsilon_n}
-\epsilon_nV\big(\frac{x_i+x_{i+1}}2\big)$.

\begin{proposition}[No limit with the weight $\epsilon_n$]\label{prop:epsweight}
Assume (H$_0$). For the weighted slice action,
$S^{\epsilon,(n)}_{\mathrm{eff}}(x,y)=\frac{m(y-x)^2}{2p^N}-p^{N-M}W_M(x,y)$ with $M=N-n$.
Hence:
(i) $S^{\epsilon,(n)}_{\mathrm{eff}}(x,x)=-p^NV(x)$ for all $n$;
(ii) for $x\ne y$ in $\Zp$ with $W^*(x,y)\ne0$ and $M\ge v_p(y-x)$,
$|p^{N-M}W_M(x,y)|_p=p^{M-N}|W^*(x,y)|_p\to\infty$ as $n\to-\infty$. No refinement limit exists,
and for fixed $a\ne0$ the propagator is inadmissible at $(x,y)$ for all sufficiently negative $n$.
\end{proposition}

\begin{proof}
Sum along the nodes of Lemma~\ref{lem:nested}, with $\epsilon_n=p^{N-M}$, and use
Proposition~\ref{prop:FP}.
\end{proof}

Theorem~\ref{thm:uniform_RG} thus depends on two things: Convention~\ref{conv:bare} avoids the
weight $\epsilon_n$, whose $p$-adic norm grows under refinement, and the estimates of
Corollary~\ref{cor:scaled_Faulhaber} control the growing number $p^M$ of terms.

\section{Variational decimation of the harmonic oscillator}\label{sec:oscillator}

For a quadratic potential on all of $\Qp$ the kinematic flow diverges under refinement
(Example~\ref{ex:nosupp}), and the appropriate rule is the variational one. In this section the
direction of the flow and the convention for the potential both change. We fix a cutoff
$\epsilon_{n_0}=p^{n_0}$ and coarsen, $\epsilon_\ell:=p^{n_0+\ell}$ with $\ell\ge0$, so that
$|\epsilon_\ell|_p\to0$; and the potential carries the weight $\epsilon$, which is $p$-adically
small in this direction. The bare action is the trapezoidal discretization
\begin{equation}\label{eq:bareHO}
  S^{(0)}(u,v)=\frac{m_0(v-u)^2}{2\epsilon_{n_0}}
  -\frac14\,m_0\omega_0^2\,\epsilon_{n_0}(u^2+v^2),
\end{equation}
with $m_0,\omega_0\in\Qp^\times$ and $g_0:=\omega_0\epsilon_{n_0}\in\Ep$. Since $g_0\in\Qp$, the
last condition means $|g_0|_p\le p^{-1}$.

\begin{definition}[Sequential elimination]\label{def:dyn}
Let $S$ be a symmetric quadratic two-point action and put $S^{(1)}_{\mathrm{part}}:=S$. For
$r=1,\dots,p-1$, if the quadratic polynomial
$y_r\mapsto S^{(r)}_{\mathrm{part}}(y_0,y_r)+S(y_r,y_{r+1})$ has non-zero leading coefficient, let
$d^*_r(y_0,y_{r+1})$ be its stationary point,
\begin{equation}\label{eq:seq_stat}
  \partial_{y_r}\Big[S^{(r)}_{\mathrm{part}}(y_0,y_r)+S(y_r,y_{r+1})\Big]=0,
\end{equation}
and put
$S^{(r+1)}_{\mathrm{part}}(y_0,y_{r+1})
:=S^{(r)}_{\mathrm{part}}\big(y_0,d^*_r\big)+S\big(d^*_r,y_{r+1}\big)$.
\end{definition}

This is Gaussian elimination for the tridiagonal linear system \eqref{eq:variational}. If all
leading coefficients are non-zero, the system has a unique solution, obtained by back substitution
$z_{p-1}=d^*_{p-1}(x,y)$, $z_{r}=d^*_{r}(x,z_{r+1})$, and $S^{(p)}_{\mathrm{part}}=\Kvar S$.

\begin{lemma}[$\sinp$ and $\cosp$ on $\Ep$]\label{lem:sin}
The series
$\sinp(z)=\sum_{j\ge0}(-1)^j\frac{z^{2j+1}}{(2j+1)!}$ and
$\cosp(z)=\sum_{j\ge0}(-1)^j\frac{z^{2j}}{(2j)!}$
converge on $\Ep$ and satisfy $\cosp^2+\sinp^2=1$ and the addition formulas there. Put
$\varkappa(z):=\frac{\sinp z}{z}-1$ for $z\ne0$.
\begin{enumerate}
\item $|\sinp z_1-\sinp z_2|_p=|z_1-z_2|_p$ on $\Ep$. The map $\sinp$ is a bijection of $\Ep$
  onto itself and of $p\Zp$ onto itself. In particular $|\sinp z|_p=|z|_p$, and $\sinp z=0$ only
  for $z=0$.
\item $|\varkappa(z)|_p\le p^{\frac2{p-1}}|z|_p^2<1$ for $z\in\Ep\setminus\{0\}$.
\item If $0<|z|_p\le p^{-1}$, then
  $|\varkappa(z)|_p=|3|_p^{-1}|z|_p^2$ and $|z\cosp z-\sinp z|_p=|3|_p^{-1}|z|_p^3$.
\end{enumerate}
\end{lemma}

\begin{proof}
Identities between power series over $\Q$ hold on $\Ep$ wherever all series converge. By
Legendre's formula $v_p((2j+1)!)\le\frac{2j}{p-1}$, so for $z\in\Ep$ and $j\ge1$,
$\big|\frac{z^{2j}}{(2j+1)!}\big|_p\le\big(|z|_p\,p^{\frac1{p-1}}\big)^{2j}
\le p^{\frac2{p-1}}|z|_p^2<1$, which gives (2).

(1) Write
$\sinp z_1-\sinp z_2=(z_1-z_2)\sum_{j\ge0}(-1)^j
\frac{z_1^{2j}+z_1^{2j-1}z_2+\dots+z_2^{2j}}{(2j+1)!}$.
The term $j=0$ is $1$, and for $j\ge1$ the $j$-th term has norm at most
$\big(\max\{|z_1|_p,|z_2|_p\}\,p^{\frac1{p-1}}\big)^{2}<1$. Hence the sum has norm $1$. For
surjectivity on $\Ep$, let $w\in\Ep\setminus\{0\}$ and $i\in\Cp$ with $i^2=-1$. The binomial
series $c:=\sum_{j\ge0}\binom{1/2}{j}(-w^2)^j$ converges, $c^2=1-w^2$ and $|c-1|_p\le|w|_p^2$.
Put $\zeta:=c+iw$; then $|\zeta-1|_p=|w|_p$ and $\zeta^{-1}=c-iw$. Since $e_p$ and the $p$-adic
logarithm are mutually inverse isometries between $\Ep$ and $1+\Ep$ \cite{Robert2000}, there is
$z\in\Ep$ with $e_p(iz)=\zeta$, and $\sinp z=(\zeta-\zeta^{-1})/(2i)=w$. Finally
$\Ep\cap\Qp=p\Zp$ is compact and $\sinp$ maps it isometrically into itself, hence onto itself.

(3) We have $\varkappa(z)=\sum_{j\ge1}(-1)^j\frac{z^{2j}}{(2j+1)!}$ and
$z\cosp z-\sinp z=\sum_{j\ge1}(-1)^j\frac{2j\,z^{2j+1}}{(2j+1)!}$. The coefficients for $j=1$ are
$-\frac16$ and $-\frac13$, of norm $|3|_p^{-1}$. For $j\ge2$ the coefficients have norm at most
$p^{v_p((2j+1)!)}$, where $v_p((2j+1)!)\le j/2$ if $p\ge5$ and $\le j$ if $p=3$. For
$|z|_p\le p^{-1}$ the ratio of the $j$-th term to the first is therefore at most
$p^{\,2-3j/2}$ if $p\ge5$ and $3^{1-j}$ if $p=3$; both are $<1$ for $j\ge2$.
\end{proof}

By Lemma~\ref{lem:sin}(1) there is a unique $\theta_0\in p\Zp\setminus\{0\}$ with
$\sinp(\theta_0/2)=g_0/2$, and $|\theta_0|_p=|g_0|_p$. Define
\begin{equation}\label{eq:renorm}
  \omega^*:=\frac{\theta_0}{\epsilon_{n_0}},\qquad
  m^*:=m_0\,\frac{\sinp\theta_0}{\theta_0},\qquad
  \theta_\ell:=\omega^*\epsilon_\ell=p^\ell\theta_0\quad(\ell\ge0).
\end{equation}

\begin{lemma}[Lattice dispersion relation]\label{lem:dispersion}
The discrete Euler--Lagrange equation of $\sum_iS^{(0)}(x_i,x_{i+1})$ is
$x_{i+1}+x_{i-1}=2\cosp(\theta_0)\,x_i$.
\end{lemma}

\begin{proof}
Differentiation with respect to $x_i$ gives
$\frac{m_0}{\epsilon_{n_0}}(2x_i-x_{i-1}-x_{i+1})-m_0\omega_0^2\epsilon_{n_0}x_i=0$, that is,
$x_{i+1}+x_{i-1}=(2-g_0^2)x_i$, and $2-g_0^2=2-4\sinp^2(\theta_0/2)=2\cosp\theta_0$.
\end{proof}

\begin{theorem}[Mehler form of the flow and its limit]\label{thm:oscillator}
Let $S^{(\ell)}:=(\Kvar)^\ell S^{(0)}$ and $\widehat S^{(\ell)}:=\epsilon_\ell S^{(\ell)}$ for
$\ell\ge0$.
\begin{enumerate}
\item For every $\ell\ge0$ and $1\le r\le p-1$ the stationary point in \eqref{eq:seq_stat}
  exists, is unique, and equals
  \begin{equation}\label{eq:dyn_centers}
    d_r^{*,(\ell)}(y_0,y_{r+1})=\frac{\sinp(\theta_\ell)\,y_0+\sinp(r\theta_\ell)\,y_{r+1}}
    {\sinp((r+1)\theta_\ell)},
  \end{equation}
  and
  \begin{equation}\label{eq:Mehler}
    S^{(\ell)}(x,y)=\frac{m^*\omega^*}{2\sinp\theta_\ell}
    \Big[\cosp\theta_\ell\,(x^2+y^2)-2xy\Big].
  \end{equation}
  As $\omega_0\to0$, $d_r^{*,(\ell)}(y_0,y_{r+1})\to\frac{y_0+r\,y_{r+1}}{r+1}$, the Dirac centre
  of HK, and $S^{(\ell)}\to\Sfree(\cdot,\cdot;\epsilon_\ell)$ with mass $m_0$.
\item For every $\ell\ge0$,
  \begin{equation}\label{eq:IR}
    \sup_{x,y\in\Zp}\Big|\widehat S^{(\ell)}(x,y)-\frac{m^*}{2}(y-x)^2\Big|_p
    =|3|_p^{-1}\,|m_0|_p\,|g_0|_p^2\,p^{-2\ell},
  \end{equation}
  and the supremum is attained at $(x,y)=(0,1)$.
\item On $\Zp^2$, $|S^{(\ell)}|_p\le|m_0|_p\,p^{n_0+\ell}$ with equality at $(0,1)$, so
  $\chi_a(S^{(\ell)})$ is not admissible on $\Zp^2$ once $|a|_p|m_0|_p\,p^{n_0+\ell}\ge\rhop$.
  If $|a|_p|m_0|_p<\rhop$, then $\chi_a(\widehat S^{(\ell)})$ is admissible on $\Zp^2$ for all
  $\ell$, and
  \[
    \sup_{x,y\in\Zp}\Big|\chi_a\big(\widehat S^{(\ell)}(x,y)\big)
    -\chi_a\Big(\frac{m^*}{2}(y-x)^2\Big)\Big|_p
    =|a|_p\,|3|_p^{-1}\,|m_0|_p\,|g_0|_p^2\,p^{-2\ell}.
  \]
\end{enumerate}
\end{theorem}

\begin{proof}
(1) Write $A_\alpha(u,v):=\frac{\kappa}{2\sinp\alpha}\big[\cosp\alpha\,(u^2+v^2)-2uv\big]$ with
$\kappa:=m^*\omega^*$. Since $\frac{\kappa}{\sinp\theta_0}=\frac{m_0}{\epsilon_{n_0}}$ and
$\cosp\theta_0=1-\frac12g_0^2$, we have $S^{(0)}=A_{\theta_0}$.

\emph{Composition.} Let $\alpha_1,\alpha_2\in\Ep$ with $\alpha_1,\alpha_2,\alpha_1+\alpha_2\ne0$,
and write $\mathsf s_i=\sinp\alpha_i$, $\mathsf c_i=\cosp\alpha_i$. Then
\[
  A_{\alpha_1}(y_0,z)+A_{\alpha_2}(z,y)
  =\frac\kappa2\,\Lambda_2z^2-\kappa\Lambda_1z+\Lambda_0,
\]
with
$\Lambda_2=\frac{\mathsf c_1}{\mathsf s_1}+\frac{\mathsf c_2}{\mathsf s_2}
=\frac{\sinp(\alpha_1+\alpha_2)}{\mathsf s_1\mathsf s_2}\ne0$,
$\Lambda_1=\frac{y_0}{\mathsf s_1}+\frac{y}{\mathsf s_2}$ and
$\Lambda_0=\frac\kappa2\big(\frac{\mathsf c_1}{\mathsf s_1}y_0^2
+\frac{\mathsf c_2}{\mathsf s_2}y^2\big)$.
The unique stationary point is
$z=\Lambda_1/\Lambda_2=\frac{\mathsf s_2y_0+\mathsf s_1y}{\sinp(\alpha_1+\alpha_2)}$, and the
stationary value is $\Lambda_0-\frac{\kappa\Lambda_1^2}{2\Lambda_2}$. Using
$\mathsf c_1\sinp(\alpha_1+\alpha_2)-\mathsf s_2=\mathsf s_1\cosp(\alpha_1+\alpha_2)$, the
coefficient of $y_0^2$ (and by symmetry of $y^2$) becomes
$\frac\kappa2\frac{\cosp(\alpha_1+\alpha_2)}{\sinp(\alpha_1+\alpha_2)}$ and the cross term
becomes $-\frac{\kappa}{\sinp(\alpha_1+\alpha_2)}y_0y$. The stationary value is therefore
$A_{\alpha_1+\alpha_2}(y_0,y)$.

\emph{Induction.} For every integer $j\ge1$ and $\ell\ge0$,
$|j\theta_\ell|_p\le|\theta_\ell|_p<\rhop$ and $j\theta_\ell\ne0$, so $\sinp(j\theta_\ell)\ne0$
by Lemma~\ref{lem:sin}(1). Assume $S^{(\ell)}=A_{\theta_\ell}$. If
$S^{(r)}_{\mathrm{part}}=A_{r\theta_\ell}$, the composition step with $\alpha_1=r\theta_\ell$,
$\alpha_2=\theta_\ell$ gives \eqref{eq:dyn_centers} and
$S^{(r+1)}_{\mathrm{part}}=A_{(r+1)\theta_\ell}$. For $r=p-1$ this is
$\Kvar S^{(\ell)}=A_{p\theta_\ell}=A_{\theta_{\ell+1}}$. The statement about $\omega_0\to0$
follows from $\sinp(j\theta)/\sinp\theta\to j$ and $m^*\to m_0$.

(2) Write $\theta=\theta_\ell$. Since $\kappa\epsilon_\ell=m^*\theta$ and
$\cosp\theta=1-2\sinp^2(\theta/2)$,
\[
  \widehat S^{(\ell)}(x,y)-\frac{m^*}{2}(y-x)^2
  =\frac{m^*}{2}\Big(\frac{\theta}{\sinp\theta}-1\Big)(y-x)^2
  -m^*\frac{\theta}{\sinp\theta}\,\sinp^2\Big(\frac{\theta}2\Big)(x^2+y^2).
\]
Here $|m^*|_p=|m_0|_p$, $|\theta/\sinp\theta|_p=1$,
$|\theta/\sinp\theta-1|_p=|\varkappa(\theta)|_p$, and $|\sinp(\theta/2)|_p=|\theta|_p$. Since
$|\theta|_p\le p^{-1}$, Lemma~\ref{lem:sin}(3) bounds the right side on $\Zp^2$ by
$|3|_p^{-1}|m_0|_p|\theta|_p^2$. At $(x,y)=(0,1)$ the left side equals
$\frac{m^*}{2}\,\frac{\theta\cosp\theta-\sinp\theta}{\sinp\theta}$, whose norm is
$|3|_p^{-1}|m_0|_p|\theta|_p^2$ by Lemma~\ref{lem:sin}(3). Finally
$|\theta_\ell|_p=p^{-\ell}|g_0|_p$.

(3) The coefficients of $\widehat S^{(\ell)}$ are $\frac{m^*\theta}{2\sinp\theta}\cosp\theta$ and
$-\frac{m^*\theta}{\sinp\theta}$, both of norm $|m_0|_p$, so $|\widehat S^{(\ell)}|_p\le|m_0|_p$
on $\Zp^2$ with equality at $(0,1)$, and $|1/\epsilon_\ell|_p=p^{n_0+\ell}$. The last formula
follows from \eqref{eq:charlip} and \eqref{eq:IR}.
\end{proof}

\begin{proposition}[The discrete classical path]\label{prop:path}
Let $K=p^\ell$. The unique solution of the equation of Lemma~\ref{lem:dispersion} with $x_0=x$
and $x_K=y$ is
\[
  x_i=\frac{\sinp((K-i)\theta_0)\,x+\sinp(i\theta_0)\,y}{\sinp(K\theta_0)}\qquad(0\le i\le K),
\]
the nodes produced by $\ell$ variational decimation steps are the points $x_{jp^{\ell-\ell'}}$
($1\le\ell'\le\ell$), and $S^{(\ell)}(x,y)=\sum_{i=0}^{K-1}S^{(0)}(x_i,x_{i+1})$.
\end{proposition}

\begin{proof}
The addition formulas show that $x_i$ satisfies the recursion. The interior equations form a
linear system whose matrix is tridiagonal with $2\cosp\theta_0$ on the diagonal and $-1$ next to
it; its determinant is $U_{K-1}(\cosp\theta_0)=\sinp(K\theta_0)/\sinp\theta_0\ne0$, where
$U_{K-1}$ is the Chebyshev polynomial of the second kind. So the fine action
$F(x_0,\dots,x_K)=\sum_iS^{(0)}(x_i,x_{i+1})$ has a unique stationary point with given endpoints.
By the composition step in the proof of Theorem~\ref{thm:oscillator}, eliminating the interior
variables of a block of $p^{\ell'}$ fine slices gives $A_{p^{\ell'}\theta_0}=S^{(\ell')}$ of the
block endpoints. The restriction of the stationary point of $F$ to a block is stationary for that
block, and its values at the block endpoints are stationary for the sum of the block actions; both
stationary points are unique. Induction on $\ell$ gives the claim.
\end{proof}

\begin{proposition}[Chebyshev recursion]\label{prop:Chebyshev}
Write \eqref{eq:Mehler} in the trapezoidal form
$S^{(\ell)}(u,v)=\frac{m_\ell(v-u)^2}{2\epsilon_\ell}
-\frac14m_\ell\omega_\ell^2\epsilon_\ell(u^2+v^2)$. Then
\[
  m_\ell=m^*\frac{\theta_\ell}{\sinp\theta_\ell},\qquad
  g_\ell:=\omega_\ell\epsilon_\ell=2\sinp\Big(\frac{\theta_\ell}2\Big),
\]
and
\begin{equation}\label{eq:Cheb}
\begin{aligned}
  g_{\ell+1}&=2(-1)^{\frac{p-1}2}\,T_p\Big(\frac{g_\ell}2\Big),\\
  \frac{m_{\ell+1}}{m_\ell}&=\frac{p\,\sinp\theta_\ell}{\sinp(p\theta_\ell)}
  =\frac{p}{U_{p-1}\big(1-\frac12g_\ell^2\big)},
\end{aligned}
\end{equation}
where $T_p$ and $U_{p-1}$ are the Chebyshev polynomials of the first and second kind; for $p=3$,
$g_{\ell+1}=3g_\ell-g_\ell^3$. Moreover $|g_\ell|_p=p^{-\ell}|g_0|_p$ and
$|m_\ell-m^*|_p=|3|_p^{-1}|m_0|_p|g_0|_p^2\,p^{-2\ell}$.
\end{proposition}

\begin{proof}
Compare coefficients in \eqref{eq:Mehler}, using $\cosp\theta_\ell=1-2\sinp^2(\theta_\ell/2)$. For
odd $p$ the identity $\sin(p\phi)=(-1)^{(p-1)/2}T_p(\sin\phi)$ holds for power series over $\Q$,
hence on $\Ep$; put $\phi=\theta_\ell/2$. The norms follow from Lemma~\ref{lem:sin}, since
$m_\ell-m^*=-m^*\varkappa(\theta_\ell)/(1+\varkappa(\theta_\ell))$.
\end{proof}

\begin{proposition}[Midpoint and displacement coordinates]\label{prop:normalization}
With $q:=\frac{u+v}2$ and $\eta:=v-u$,
\begin{equation}\label{eq:sd}
  \widehat S^{(\ell)}(u,v)=-\frac{m_\ell g_\ell^2}{2}\,q^2
  +\frac{m_\ell}{2}\Big(1-\frac{g_\ell^2}{4}\Big)\eta^2 ,
\end{equation}
and $\widehat S^{(\ell+1)}=(\Dp\circ\Kvar)\,\widehat S^{(\ell)}$. The coefficient of $q^2$ is
multiplied at each step by $p^2\,(1+O(g_\ell^2))$, and the coefficient of $\eta^2$ tends to
$\frac12m^*$.
\end{proposition}

\begin{proof}
Substitute $u^2+v^2=2q^2+\frac12\eta^2$ in the trapezoidal form. The identity for
$\widehat S^{(\ell+1)}$ holds because $\Kvar$ commutes with multiplication by non-zero scalars and
$\epsilon_{\ell+1}=p\,\epsilon_\ell$. The last statement follows from \eqref{eq:Cheb}.
\end{proof}

\begin{remark}[Real and $p$-adic direction of the flow]\label{rem:irrelevant}
At $g=0$ the recursion \eqref{eq:Cheb} reads $g_{\ell+1}=p\,g_\ell+O(g_\ell^3)$. The multiplier is
the rational prime $p$, and $|p|_\infty|p|_p=1$. Over $\R$ the fixed point $g=0$ is repelling:
orbits stay bounded if $|g_0|_\infty\le2$, where $g_\ell=2\sin(p^\ell\theta_0/2)$, and tend to
infinity if $|g_0|_\infty>2$. Over $\Qp$ the fixed point is attracting, $\theta_\ell$ contracts
inside $\Ep$, no caustic occurs, and $\widehat S^{(\ell)}$ tends to the free line with error of
order $|g_\ell|_p^2$. In $p$-adic terms the limit $\ell\to\infty$ is therefore a small-angle limit,
and the mass renormalization $m^*/m_0=\sinp(\theta_0)/\theta_0$ records how far the initial angle
is from zero. The obstruction to a refinement limit in Proposition~\ref{prop:noUV} is a separate
fact, due to the finite radius of $\Ep$.
\end{remark}

\begin{figure}[t]
\centering
\begin{tikzpicture}[scale=0.92]
  \draw[->] (-0.5,0) -- (6.5,0) node[right]{$g$};
  \draw[->] (0,-0.5) -- (0,4.0) node[above]{$m$};
  \fill (0,3.15) circle (2.2pt);
  \node[anchor=west] at (0.15,3.42) {\small $(0,m^*)$};
  \draw[thick,->] (5.4,0.55) .. controls (3.8,1.7) and (2.2,2.6) .. (0.5,3.1);
  \fill (5.4,0.55) circle (1.6pt) node[anchor=west]{$(g_0,m_0)$};
  \fill (2.8,2.05) circle (1.6pt) node[anchor=south west]{$(g_1,m_1)$};
  \fill (1.2,2.85) circle (1.6pt) node[anchor=south west]{$(g_2,m_2)$};
  \node[anchor=west] at (3.9,1.25) {\small $g_{\ell+1}=3g_\ell-g_\ell^3$};
  \draw[->,thick] (2.0,0.25) -- (0.4,0.25)
    node[above,midway]{\small $\Qp$: $|g_\ell|_p=p^{-\ell}|g_0|_p$};
  \draw[->,thick] (0.4,-0.3) -- (2.0,-0.3)
    node[below,midway]{\small $\R$: $g=0$ repelling};
\end{tikzpicture}
\caption{Schematic flow of the oscillator couplings $(g_\ell,m_\ell)$ in the $p$-adic norm, for
$p=3$. The coupling contracts and the mass converges to $m^*$; in the real norm the fixed point
$g=0$ is repelling.}
\label{fig:RG}
\end{figure}
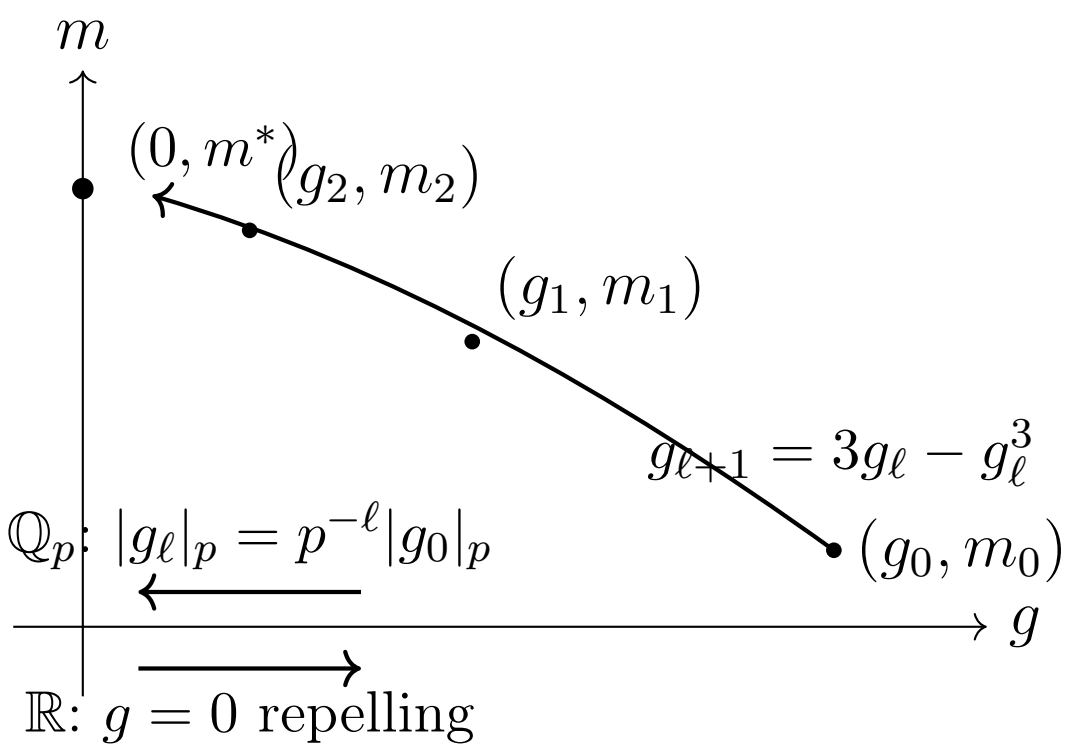

\begin{proposition}[No refinement limit at fixed frequency]\label{prop:noUV}
Fix $m^*,\omega^*\in\Qp^\times$. A bare action \eqref{eq:bareHO} at cutoff $n_0$ with
$g_0\in\Ep$ whose flow is \eqref{eq:Mehler} with parameters $(m^*,\omega^*)$ exists if and only if
$|\omega^*|_p\,p^{-n_0}<\rhop$, that is, $n_0\ge1-v_p(\omega^*)$. In particular $n_0\to-\infty$
is impossible.
\end{proposition}

\begin{proof}
The angle $\theta_0=\omega^*\epsilon_{n_0}$ must lie in $\Ep$, and
$|\epsilon_{n_0}|_p=p^{-n_0}$. Conversely, if $\theta_0\in\Ep\setminus\{0\}$, put
$\omega_0:=2\sinp(\theta_0/2)/\epsilon_{n_0}$ and $m_0:=m^*\theta_0/\sinp\theta_0$.
\end{proof}

\begin{remark}[Other discretizations]\label{rem:universality}
A symmetric quadratic action $\frac A2(u^2+v^2)-Buv$ with $A,B\in\Cp$, $B\ne0$, is of the form
\eqref{eq:Mehler} if and only if $A/B=\cosp\theta_0$ for some $\theta_0\in\Ep\setminus\{0\}$,
that is, $0<|1-A/B|_p<p^{-\frac2{p-1}}$. For the midpoint discretization
$-\frac12m_0\omega_0^2\epsilon_{n_0}\big(\frac{u+v}2\big)^2$ of the potential one finds
$\cosp\theta_0=\frac{1-g_0^2/4}{1+g_0^2/4}$. Different discretizations flow to the same free line
and differ in the map $(m_0,\omega_0)\mapsto(m^*,\omega^*)$.
\end{remark}

\begin{remark}[No fluctuation prefactor]\label{rem:realcase}
Over $\R$, Gaussian integration over the slice variables multiplies the exponential of the Mehler
action by the Van Vleck factor $(m\omega/(2\pi i\hbar\sin\omega T))^{1/2}$, which is singular at
the caustics $\sin\omega T=0$. Here no prefactor arises, because point evaluation integrates out
no fluctuations, and the action itself has no singularity because $\sinp$ has no zero in
$\Ep\setminus\{0\}$.
\end{remark}

\section{The linearized map at the free line}\label{sec:spectrum}

By Proposition~\ref{prop:tangent} the linearization of $\Dp\circ\Kvar$ at a free action is
$\sT=p\,\Kkin$. Since $\Kkin$ shifts the polydisc (Proposition~\ref{prop:continuous}), no single
Tate algebra is $\sT$-invariant, and we study $\sT$ on the polynomial ring $\Cp[u,v]$, graded by
degree. Put $q=\frac{u+v}2$, $\eta=v-u$, and for $k\ge0$
\[
\begin{aligned}
  H_k(u,v):=\frac{v^{k+1}-u^{k+1}}{(k+1)(v-u)}
  &=\int_0^1\big(u+\lambda(v-u)\big)^k\dd\lambda\\
  &=\sum_{\substack{0\le j\le k\\ j\ \mathrm{even}}}\binom kj\frac{1}{(j+1)2^j}\,q^{k-j}\eta^{j}.
\end{aligned}
\]

\begin{proposition}[Eigenbasis of $\sT$]\label{prop:spectrum}
Let $\beta_r:=\frac{2r+1-p}{2p}$ for $0\le r<p$, so that the $r$-th kinematic slice has midpoint
$q+\beta_r\eta$ and displacement $\eta/p$.
\begin{enumerate}
\item For all integers $k,b\ge0$,
  \begin{equation}\label{eq:eigen}
    \sT\big[\eta^bH_k\big]=p^{2-b}\,\eta^bH_k .
  \end{equation}
  The polynomials $\eta^bH_k$ with $k+b=D$ form a basis of the homogeneous polynomials of degree
  $D$, and those with $b$ even form a basis of the symmetric ones. Hence $\sT$ is diagonalizable
  on $\Cp[u,v]$, and on symmetric homogeneous polynomials of degree $D$ its eigenvalues are
  $p^{2-b}$, $b\in\{0,2,\dots,2\lfloor D/2\rfloor\}$, each simple.
\item In the monomial basis,
  \begin{equation}\label{eq:triangular}
    \sT\big[q^k\eta^b\big]
    =p^{2-b}q^k\eta^b+p^{1-b}\sum_{\substack{2\le j\le k\\ j\ \mathrm{even}}}\binom kj
     \Big(\sum_{r=0}^{p-1}\beta_r^{\,j}\Big)q^{k-j}\eta^{b+j}.
  \end{equation}
\item In degree $2$,
  \begin{equation}\label{eq:offdiag}
  \begin{aligned}
    \sum_{r=0}^{p-1}\beta_r^2&=\frac{p^2-1}{12p},\qquad
    \sT[q^2]=p^2q^2+\frac{p^2-1}{12}\,\eta^2,\\
    \Big|\frac{p^2-1}{12}\Big|_p&=|3|_p^{-1},
  \end{aligned}
  \end{equation}
  and the eigenvector for $p^2$ is $H_2=q^2+\frac1{12}\eta^2=\frac13(u^2+uv+v^2)$.
\end{enumerate}
\end{proposition}

\begin{proof}
(1) With $z_r=u+\frac rp(v-u)$, the sum telescopes:
\[
  \Kkin\big[\eta^bH_k\big]
  =\sum_{r=0}^{p-1}\Big(\frac\eta p\Big)^{b-1}\frac{z_{r+1}^{k+1}-z_r^{k+1}}{k+1}
  =p^{1-b}\,\eta^{b-1}\,\frac{v^{k+1}-u^{k+1}}{k+1}=p^{1-b}\,\eta^bH_k ,
\]
an identity of polynomials (for $b=0$ it is first read for $\eta\ne0$). The expansion of $H_k$
shows that the passage from $\{q^k\eta^b\}$ to $\{\eta^bH_k\}$ is unitriangular, and $H_k$ is
symmetric while $\eta^b$ has parity $(-1)^b$ under $u\leftrightarrow v$.

(2) By the binomial theorem,
$\sT[q^k\eta^b]=p\sum_r(q+\beta_r\eta)^k(\eta/p)^b
=p^{1-b}\sum_{j}\binom kj\big(\sum_r\beta_r^j\big)q^{k-j}\eta^{b+j}$. The term $j=0$ is
$p^{2-b}q^k\eta^b$, and $\beta_{p-1-r}=-\beta_r$ removes the odd $j$.

(3) $\sum_{r=0}^{p-1}(2r+1-p)^2=\frac{p(p^2-1)}3$, so
$\sum_r\beta_r^2=\frac{p^2-1}{12p}$, and \eqref{eq:triangular} with $k=2$, $b=0$ gives the formula
for $\sT[q^2]$. For $p\ge5$ the number $\frac{p^2-1}{12}$ is a $p$-adic unit; for $p=3$ it is
$\frac23$.
\end{proof}

\begin{remark}\label{rem:telescoping}
The telescoping argument does not use that $H_k$ is a polynomial: for any $F:\Qp\to\Cp$ and
$b\ge1$, the action $(v-u)^{b-1}(F(v)-F(u))$ is an eigenvector of $\Kkin$ with eigenvalue
$p^{1-b}$. For $b=1$ these are the boundary terms $F(v)-F(u)$, which $\Kkin$ leaves unchanged.
\end{remark}

\begin{remark}[Consequences]\label{rem:center_stable}
(i) \emph{Classification.} In the sense of Definition~\ref{def:relevance}, the directions with
$b=0$ (potentials averaged along the segment) are $p$-adically irrelevant, the direction $b=2$,
$k=0$ is the marginal kinetic term, the directions with $b=2$ and $k\ge1$ (position-dependent
mass) are marginal, and the directions with $b\ge4$ are $p$-adically relevant. Since the
eigenvalues are the rational numbers $p^{2-b}$ and $|p^{2-b}|_\infty|p^{2-b}|_p=1$, the
classification over $\R$ is the reverse. The free line is therefore not attracting on all
polynomial actions in either norm; $p$-adically it attracts within the sector $b\le2$, which
contains the quadratic actions of Section~\ref{sec:oscillator}.

(ii) \emph{The monomial $q^2$ is not an eigenvector.} The $p$-adically irrelevant direction of
degree $2$ is $H_2$, not $q^2$: by \eqref{eq:offdiag}, $\sT[q^2]$ has a component of norm
$|3|_p^{-1}\ge1$ along the marginal direction $\eta^2$. This is the mechanism behind
Example~\ref{ex:nosupp}.

(iii) \emph{Mass renormalization at first order.} Write
$\widehat S^{(\ell)}=\frac{m_\ell}2\eta^2-U_\ell$ with
$U_\ell=\frac{m_\ell g_\ell^2}{2}\big(q^2+\frac14\eta^2\big)$, by \eqref{eq:sd}. Then
$\sT U_\ell=\frac{m_\ell g_\ell^2}2\big(p^2q^2+\frac{p^2+2}{12}\eta^2\big)$, so the linearized
step predicts the coefficient $\frac{m_\ell}2\big(1-\frac{p^2+2}{12}g_\ell^2\big)$ of $\eta^2$ at
step $\ell+1$. The exact value is $\frac{m_{\ell+1}}2\big(1-\frac14g_{\ell+1}^2\big)$, and
\eqref{eq:Cheb} gives $m_{\ell+1}/m_\ell=1+\frac{p^2-1}6g_\ell^2+O(g_\ell^4)$ and
$g_{\ell+1}^2=p^2g_\ell^2+O(g_\ell^4)$, hence the same coefficient up to $O(g_\ell^4)$. This
agrees with the quadratic remainder in \eqref{eq:tangent}. The mass shift $m_\ell\to m^*$ is thus
produced, at leading order, by the off-diagonal entry in \eqref{eq:offdiag}.
\end{remark}

\section{Concluding remarks and open problems}\label{sec:discussion}

The finite-partition path integral of HK is, at the level of actions, an exact block decimation
along a chain. The arithmetic relation $|p|_\infty|p|_p=1$ then reverses the roles that refinement
and coarsening play over $\R$: refinement converges for potentials cut off outside $\Zp$ when the
potential carries no factor $\epsilon_n$, and coarsening converges for the oscillator when it
does. We list some problems that the present results leave open.

\begin{enumerate}
\item \emph{Non-atomic blocking kernels.} Replace the Dirac measures in \eqref{eq:nested} by
  kernels that integrate over a neighbourhood of the nodes. Since no non-zero bounded measure is
  translation invariant, natural candidates are either unbounded distributions, such as the
  Volkenborn distribution or Khrennikov's Gaussian distributions \cite{Khrennikov1994book,AK1996},
  or bounded measures that are not translation invariant, such as the measure
  $\mu_{-1}(a+p^L\Zp)=(-1)^a$ ($0\le a<p^L$) underlying the fermionic $p$-adic integral
  \cite{Kim2007,KimHu2012}. For each choice one should determine whether a fluctuation prefactor
  arises and compare the refinement asymptotics with Proposition~\ref{prop:Volkenborn}.
\item \emph{Larger supports.} Extend Theorem~\ref{thm:uniform_RG} to potentials supported on
  $p^{-r}\Zp$ or on general compact-open sets. Example~\ref{ex:nosupp} suggests that an explicit
  subtraction along the marginal $\eta^2$ direction is then required.
\item \emph{Beyond first order.} For potentials as in (H), compare the kinematic limit $W^*$ with
  the variational flow $\Kvar$, and describe the stable set of the free line under
  $\Dp\circ\Kvar$ in the presence of the $p$-adically relevant higher-gradient terms ($b\ge4$).
\item \emph{Time dependence and composition.} Clarify the role of identifying $p^N$ with
  $p^{-1}|t|_p^{-1}$ in HK, and decide whether any prescription of the present type admits a
  composition law  in the sense of Chapman--Kolmogorov in which the intermediate position is integrated.
\item \emph{Higher dimensions.} Extend block decimation to $\Qp^d$ ($d\ge2$) and to $\Cp$, where
  Zelenov's linear ordering of segments is absent.
\end{enumerate}

\appendix
\section{Worked examples for Section~\ref{sec:perfect}}\label{app:A}

In this appendix $p=3$ unless stated otherwise, and $s=v_p(y-x)$ as in
Section~\ref{sec:perfect}. All quantities are rational numbers.

\begin{example}[The bound of Lemma~\ref{lem:Faulhaber} is attained]\label{ex:A_Faulhaber}
For $L=2$ and $P_k:=\sum_{i=0}^{8}i^k$:
\begin{center}
\begin{tabular}{c|ccccc}
$k$ & $0$ & $1$ & $2$ & $3$ & $4$\\ \hline
$P_k$ & $9$ & $36$ & $204=2^2\cdot3\cdot17$ & $1296=2^4\cdot3^4$ & $8772=2^2\cdot3\cdot17\cdot43$\\
$v_3(P_k)$ & $2$ & $2$ & $1$ & $4$ & $1$
\end{tabular}
\end{center}
The bound $v_3(P_k)\ge L-1=1$ for $k\ge1$ is attained at $k=2$ and $k=4$.
\end{example}

\begin{example}[Stabilization]\label{ex:A_lock}
Let $V(z)=z^2\mathbf 1_{\Z_3}(z)$, so $\nT{V}=1$, and $x=0$, $y=3$, $s=1$. The midpoints at
step $M$ are $\frac{(2i+1)3^{1-M}}2$, which lie in $\Z_3$ if and only if $3^{M-1}\mid(2i+1)$:
\begin{center}
\begin{tabular}{c|l|c}
$M$ & midpoints in $\Z_3$ & $W_M(0,3)$\\ \hline
$0$ & $\tfrac32$ & $\tfrac94$\\
$1$ & $\tfrac12,\,\tfrac32,\,\tfrac52$ & $\tfrac{35}4$\\
$2$ & $\tfrac36,\,\tfrac96,\,\tfrac{15}6$ \ (of $9$) & $\tfrac{35}4$\\
$3$ & $\tfrac9{18},\,\tfrac{27}{18},\,\tfrac{45}{18}$ \ (of $27$) & $\tfrac{35}4$
\end{tabular}
\end{center}
The interaction is constant from $M=s=1$ on, as in Theorem~\ref{thm:uniform_RG}(1). With
$z_j=j$, $\sum_{j=0}^{2}W_0(j,j+1)=\frac{1+9+25}4=W_1(0,3)$, as in Proposition~\ref{prop:FP}(1),
and $|W^*(0,3)|_3=1=p^{1-s}$.
\end{example}

\begin{example}[The rate is attained off the diagonal]\label{ex:A_rate}
Let $V(z)=z^2\mathbf 1_{\Z_3}(z)$, $x=0$, $y=27$, $s=3$. For $M\le3$ all $K=3^M$ midpoints lie
in $\Z_3$, and
\[
  W_M(0,27)=\frac{729\,(4\cdot9^M-1)}{12\cdot3^M},\qquad W^*(0,27)=W_3(0,27)=\frac{26235}{4}.
\]
\begin{center}
\begin{tabular}{c|c|c|c|c}
$M$ & $W_M(0,27)$ & $v_3(W_M)$ & $v_3(W_M-W^*)$ & lower bound from \eqref{eq:rate}\\ \hline
$0$ & $729/4$   & $6$ & $2$ & $0$\\
$1$ & $2835/4$  & $4$ & $2$ & $1$\\
$2$ & $8721/4$  & $3$ & $2$ & $2$\\
$3$ & $26235/4$ & $2$ & $\infty$ & $3$
\end{tabular}
\end{center}
Here $W_M-W^*=-\frac{25506}4,-\frac{23400}4,-\frac{17514}4$ for $M=0,1,2$. The bound
\eqref{eq:rate} is an equality at $M=s-1=2$.
\end{example}

\begin{example}[Volkenborn asymptotics]\label{ex:A_Volk}
Let $p$ be any odd prime, $V(z)=z^2\mathbf 1_{\Zp}(z)$, $x=0$, $y=p^s$ with $s\ge1$.
Then $\varrho=1$, $g_{0,1}(z)=(z+\frac12)^2$ and
$\int_{\Zp}g_{0,1}\dd\mu_{\mathrm{Vol}}=B_2+B_1+\frac14=-\frac1{12}$. With $K=p^s$,
\[
  W^*(0,p^s)=\frac{K(4K^2-1)}{12},\qquad
  W^*(0,p^s)-p^s\!\int_{\Zp}g_{0,1}\dd\mu_{\mathrm{Vol}}=\frac{K^3}{3}.
\]
The remainder has norm $|3|_p^{-1}p^{-3s}$, within the bound $p^{1-2s}$ of
Proposition~\ref{prop:Volkenborn}. For $p=3$:
\begin{center}
\begin{tabular}{c|c|c|c|c}
$s$ & $W^*(0,3^s)$ & $3^s\int g_{0,1}\dd\mu_{\mathrm{Vol}}$ & difference
 & $v_3$ (bound: $\ge2s-1$)\\ \hline
$1$ & $35/4$  & $-1/4$ & $9$   & $2$\\
$2$ & $969/4$ & $-3/4$ & $243$ & $5$
\end{tabular}
\end{center}
\end{example}

\begin{example}[Divergence without compact support]\label{ex:A_nosupp}
For $V(z)=z^2$ on all of $\Q_3$, Example~\ref{ex:nosupp} gives
$W_M(0,1)=\frac{4\cdot9^M-1}{12\cdot3^M}$, that is, $\frac14,\frac{35}{36},\frac{323}{108},
\frac{2915}{324}$ for $M=0,\dots,3$, and $|W_M(0,1)|_3=3^{M+1}$ for $M\ge1$.
\end{example}

\begin{example}[Divergence with the weight $\epsilon_n$]\label{ex:A_eps}
Let $V=\mathbf 1_{\Z_3}$, $N=0$, $x=0$, $y=1$, so that $s=0$ and $W_M(0,1)=1$. Under
Convention~\ref{conv:bare}, $S^{(n)}_{\mathrm{eff}}(0,1)=\frac m2-1$ for all $n\le0$. With the
weight $\epsilon_n$ of Proposition~\ref{prop:epsweight},
$S^{\epsilon,(n)}_{\mathrm{eff}}(0,1)=\frac m2-3^{-M}$, and $|3^{-M}|_3=3^M\to\infty$.
\end{example}

\section{Worked examples for Section~\ref{sec:oscillator}}\label{app:B}

Let $p=3$, $n_0=0$, $m_0=1$ and $\omega_0=3$. Then $g_0=3$, $|g_0|_3=3^{-1}$, and
$\mathsf c_0:=\cosp\theta_0=1-\frac12g_0^2=-\frac72$. From the series of the inverse of $\sinp$
one finds $\theta_0\equiv3+\frac98\pmod{3^6\Z_3}$, and
\[
  S^{(0)}(u,v)=\tfrac12(v-u)^2-\tfrac94(u^2+v^2)=-\tfrac74(u^2+v^2)-uv .
\]

\begin{example}[One decimation step]\label{ex:B_step}
\emph{Step $r=1$.} The equation
$\partial_{y_1}[S^{(0)}(y_0,y_1)+S^{(0)}(y_1,y_2)]=-7y_1-y_0-y_2=0$ gives
\[
  d_1^{*,(0)}(y_0,y_2)=-\frac{y_0+y_2}7=\frac{\sinp\theta_0\,(y_0+y_2)}{\sinp(2\theta_0)},
\]
since $\sinp(2\theta_0)/\sinp\theta_0=2\mathsf c_0=-7$. Substitution gives
$S^{(2)}_{\mathrm{part}}(y_0,y_2)=-\frac{47}{28}(y_0^2+y_2^2)+\frac17y_0y_2$, which is the Mehler
form with angle $2\theta_0$: $\cosp(2\theta_0)=2\mathsf c_0^2-1=\frac{47}2$ and
$m^*\omega^*/\sinp(2\theta_0)=-\frac17$.

\emph{Step $r=2$.} The equation
$\partial_{y_2}[S^{(2)}_{\mathrm{part}}(y_0,y_2)+S^{(0)}(y_2,y_3)]
=-\frac{48}7y_2+\frac{y_0}7-y_3=0$ gives
\[
  d_2^{*,(0)}(y_0,y_3)=\frac{y_0-7y_3}{48}
  =\frac{\sinp\theta_0\,y_0+\sinp(2\theta_0)\,y_3}{\sinp(3\theta_0)},
\]
since $\sinp(3\theta_0)/\sinp\theta_0=4\mathsf c_0^2-1=48$, and
\[
  S^{(1)}(x,y)=-\frac{161}{96}(x^2+y^2)-\frac1{48}xy
  =\frac1{96}\Big[\cosp(3\theta_0)(x^2+y^2)-2xy\Big],
\]
in agreement with \eqref{eq:Mehler}, because $\cosp(3\theta_0)=4\mathsf c_0^3-3\mathsf c_0=-161$.

\emph{Classical path.} Proposition~\ref{prop:path} with $K=3$ gives $x_1=\frac{-7x+y}{48}$ and
$x_2=\frac{x-7y}{48}$. These agree with $x_2=d_2^{*,(0)}(x,y)$, $x_1=d_1^{*,(0)}(x,x_2)$ and
solve $x_0+x_2=-7x_1$, $x_1+x_3=-7x_2$.
\end{example}

\begin{remark}\label{rem:rational}
Once $\mathsf c_0=-\frac72$ is given, the elimination in Example~\ref{ex:B_step} takes place over
$\Q$. The $p$-adic analysis enters in two places: the denominators
$\sinp(k\theta_0)/\sinp\theta_0=U_{k-1}(\mathsf c_0)$ do not vanish for $g_0\in\Ep$, and an angle
$\theta_0\in\Ep$ with $\cosp\theta_0=-\frac72$ exists. Over $\R$, $|\mathsf c_0|>1$ means
$\omega_0\epsilon_{n_0}=3>2$, the unstable regime of the discretized oscillator.
\end{remark}

\begin{example}[Couplings and mass]\label{ex:B_traj}
For $p=3$, \eqref{eq:Cheb} reads $g_{\ell+1}=3g_\ell-g_\ell^3$ and
$m_{\ell+1}/m_\ell=3/(4\mathsf c_\ell^2-1)$ with $\mathsf c_\ell=1-\frac12g_\ell^2$:
\begin{center}
\begin{tabular}{c|c|c|c|c|c}
$\ell$ & $g_\ell$ & $v_3(g_\ell)$ & $\mathsf c_\ell$ & $m_\ell$ & $v_3(m_\ell-m_{\ell-1})$\\ \hline
$0$ & $3$     & $1$ & $-7/2$ & $1$ & --\\
$1$ & $-18$   & $2$ & $-161$ & $1/16$ & $1$\\
$2$ & $5778$  & $3$ & $-16692641$ & $1/552976$ & $3$
\end{tabular}
\end{center}
One checks $1-\frac{5778^2}2=-16692641=4(-161)^3-3(-161)$, and $-\frac{m_1}{\epsilon_1}xy
=-\frac1{48}xy$ as in Example~\ref{ex:B_step}. Here $|g_\ell|_\infty\to\infty$ because
$|g_0|_\infty>2$, while $|g_\ell|_3=3^{-1-\ell}$. By Proposition~\ref{prop:Chebyshev},
$|m_\ell-m^*|_3=3^{-1-2\ell}$, so $m^*\equiv m_2\equiv103\pmod{3^5}$, consistent with
$m_1\equiv22\pmod{3^3}$ and $m_0\equiv1\pmod3$.
\end{example}

\begin{example}[The limit formula for $p=3$]\label{ex:B_IR}
At $(x,y)=(0,1)$ and $\ell=1$,
$\widehat S^{(1)}(0,1)-\frac{m^*}2=\frac{m_1-m^*}2-\frac{m_1g_1^2}4$. The first term has norm
$3^{-3}$ and the second has norm $|324/64|_3=3^{-4}$, so the difference has norm $3^{-3}$, which
is $|3|_3^{-1}|m_0|_3|g_0|_3^2\,3^{-2}$ as in \eqref{eq:IR}.
\end{example}

\begin{example}[The case $p=5$]\label{ex:B_p5}
Let $p=5$, $\epsilon_{n_0}=1$, $m_0=1$, $\omega_0=5$, so $g_0=5$ and $\mathsf c_0=-\frac{23}2$.
Then $g_1=2T_5(g_0/2)=g_0^5-5g_0^3+5g_0=2525=5^2\cdot101$ and
$m_1=\frac{5}{U_4(\mathsf c_0)}=\frac{5}{16\mathsf c_0^4-12\mathsf c_0^2+1}=\frac{5}{278255}
=\frac{1}{55651}$. Since $m_1-m_0=-\frac{55650}{55651}$ has $5$-adic valuation $2$ and
$|m_1-m^*|_5=5^{-4}$, we get $|m_0-m^*|_5=5^{-2}=|g_0|_5^2$, in agreement with
Proposition~\ref{prop:Chebyshev} for $|3|_5=1$.
\end{example}

\begin{remark}[Where the real caustics lie $p$-adically]\label{rem:caustic}
A caustic after $k$ steps means $U_{k-1}(\mathsf c)=0$ with $\mathsf c=1-\frac12g^2$. The roots
are $\mathsf c=\frac12(\zeta+\zeta^{-1})$ with $\zeta$ a $2k$-th root of unity, $\zeta\ne\pm1$,
so that $g^2=2-\zeta-\zeta^{-1}=-(1-\zeta)^2/\zeta$; over $\R$ this is $g^2=4\sin^2(\pi j/2k)$.
In $\Cp$, $|g|_p=|1-\zeta|_p$, which equals $1$ unless the order of $\zeta$ is a power $p^e$ of
$p$, in which case it equals $p^{-1/(p^{e-1}(p-1))}\ge\rhop$. Hence $|g|_p\ge\rhop$ for every
root, with equality exactly when $\zeta$ has order $p$, and the condition $g_0\in\Ep$ excludes
all of them. For $p=3$ and $k=3$: $U_2(\mathsf c)=4\mathsf c^2-1$ has the roots
$\mathsf c=-\frac12$, where $g^2=3$ and $|g|_3=3^{-1/2}=\rhop$, and $\mathsf c=\frac12$, where
$g^2=1$.
\end{remark}


\end{document}